\documentclass[final,3p,times]{elsarticle}

\usepackage{amssymb}

\usepackage{newtxtext}
\usepackage[varvw]{newtxmath}

\usepackage{graphicx}

\usepackage{caption}

\usepackage{subcaption}

\usepackage{cancel}

\usepackage{multirow}

\usepackage{multicol}

\usepackage{fontawesome}

\usepackage{wasysym}

\usepackage{bm}

\usepackage{orcidlink}

\journal{Tribology International}

\begin{document}

\begin{frontmatter}

\title{Sliding Speed Influences Electrovibration-Induced Finger Friction Dynamics on Touchscreens}

\author[label1]{Jagan K. Balasubramanian\orcidlink{0000-0002-8990-9629}} 

\author[label2]{Daan M. Pool\orcidlink{0000-0001-9535-2639}} 

\author[label1]{Yasemin Vardar\orcidlink{0000-0003-2156-1504}\footnote{Corresponding author, 
y.vardar@tudelft.nl}} 

\affiliation[label1]{organization={Dept. of Cognitive Robotics, Mechanical Engineering, TU Delft},
            addressline={Mekelweg 5}, 
            city={Delft},
            postcode={2628 CD}, 
            country={The Netherlands}}

\affiliation[label2]{organization={Dept. of Control \& Operations, Aerospace Engineering, TU Delft},
             addressline={Kluyverweg 1}, 
            city={Delft},
            postcode={2629 HS}, 
            country={The Netherlands}}




\begin{abstract}
Electrovibration technology enables tactile texture rendering on capacitive touchscreens by modulating friction between the finger and the screen through electrostatic attraction forces, generated by applying an alternating voltage signal to the screen. Accurate signal calibration is essential for robust texture rendering but remains challenging due to variations in sliding speed, applied force, and individual skin mechanics, all of which unpredictably affect frictional behavior. Here, we investigate how exploration conditions affect electrovibration-induced finger friction on touchscreens and the role of skin mechanics in this process. Ten participants slid their index fingers across an electrovibration-enabled touchscreen at five sliding speeds ($20\sim100$~mm/s) and applied force levels ($0.2\sim0.6$~N). Contact forces and skin accelerations were measured while amplitude modulated voltage signals spanning the tactile frequency range were applied to the screen. We modeled the finger-touchscreen friction response as a first-order system and the skin mechanics as a mass-spring-damper system. Results showed that sliding speed influenced the friction response's cutoff frequency, along with the estimated finger moving mass and stiffness. For every 1~mm/s increase in speed, the cutoff frequency, the finger moving mass, and stiffness increased by 13.8~Hz, $3.23\times 10^{-5}$~kg, and 4.04~N/m, respectively. Correlation analysis revealed that finger stiffness had a greater impact on the cutoff frequency than moving mass. Notably, we observed a substantial inter-participant variability in both finger-display interaction and skin mechanics parameters. Finally, we developed a speed-dependent friction model to support consistent and perceptually stable electrovibration-based haptic feedback across varying user conditions.
\end{abstract}

\begin{graphicalabstract}
\includegraphics[width=\linewidth]{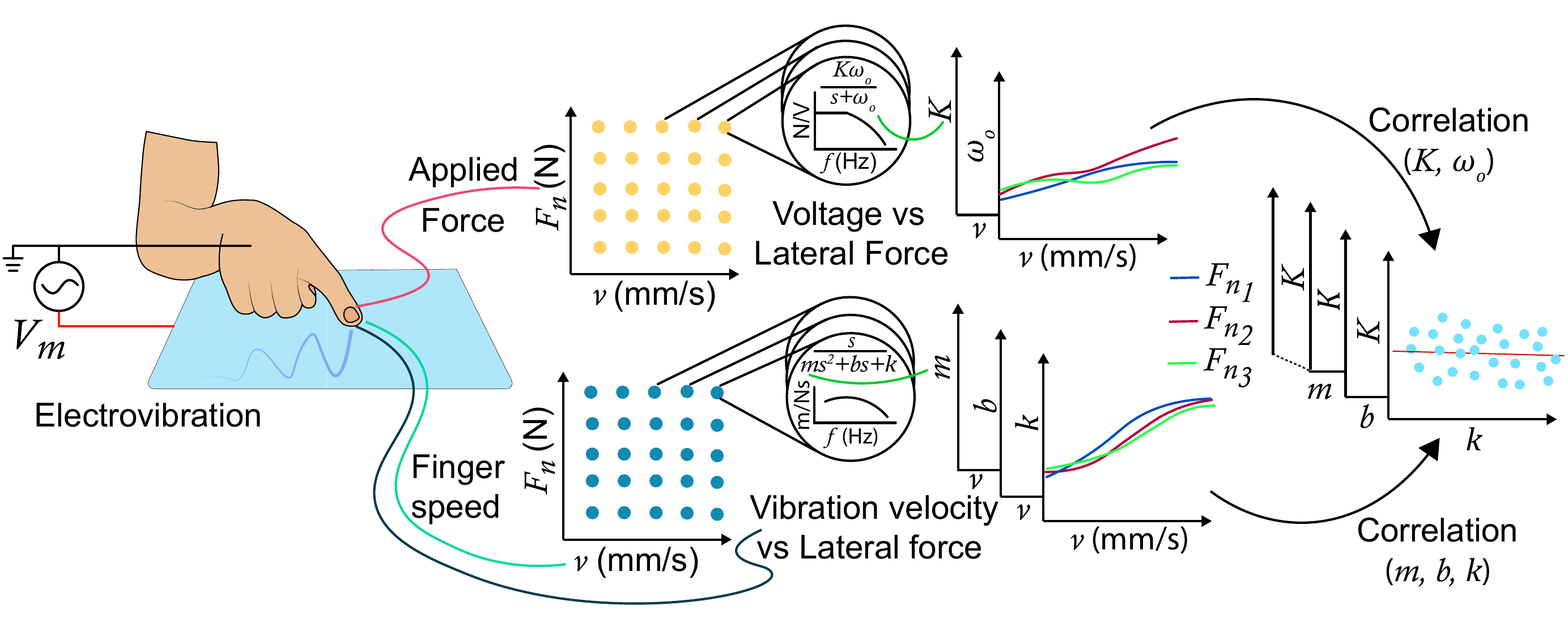}
\end{graphicalabstract}

\begin{highlights}
\item Sliding speed impacts electrovibration-induced finger friction bandwidth
\item Applied force does not affect electrovibration-induced finger friction dynamics
\item Skin stiffness influences finger-electrovibration bandwidth more than mass does 
\item Electrovibration-induced finger friction dynamics vary across participants
\item A model is proposed to compensate for speed-based variations in electrovibration friction
\end{highlights}


\end{frontmatter}

\section{Introduction}\label{sec:introduction}

Imagine a touchscreen that does not just sense your touch but lets you feel the textures -- whether it is the roughness of tree bark or the smoothness of silk -- beneath your fingers. The field of surface haptics aims to generate artificial tactile sensations, such as mimicking the touch experience of interacting with physical textures on flat surfaces, by modulating friction or vibrations between the user's finger and a flat surface~\cite{basdogan2020review}. Recreating the experience of touching physical surfaces -- known as data-based texture rendering -- involves recording the contact forces or skin vibrations when a finger interacts with physical textures. Then, these finger-surface interaction recordings serve as input for surface haptic devices equipped with actuators, which convert electrical signals into friction or vibration outputs. 

Electrovibration is one such technology that can modulate friction between a touchscreen and a user's finger via electrostatic forces~\cite{mallinckrodt1953perception}. When a finger interacts with the touchscreen, it functions as a parallel plate capacitor, where the screen's non-conductive coating, the outermost layer of the skin (stratum corneum), and an air gap between the skin and the touchscreen (caused by the finger ridges and touchscreen's rough surface) act as the dielectric layers~\cite{shultz2015surface,shultz2018electrical,vodlak2016multi}. The touchscreen's conductive layer beneath the non-conductive coating and the skin's inner tissues serve as two parallel conductors. Applying an alternating high-voltage signal to the conductive layer of the touchscreen generates an electrostatic force between the finger and the touchscreen, pulling the skin toward the surface in response to the input voltage signal. Although this force is very weak when the finger is static~\cite{vardar2021finger}, it modulates friction during finger movement by repeatedly attracting and releasing the skin, thereby creating the sensation of texture on an otherwise smooth glass surface. By using recorded contact forces from real-world finger-surface interactions as input signals, electrovibration displays can replicate virtual texture sensations that closely match their physical counterparts~\cite{fiedler2019novel,grigorii2021data}. Electrovibration outperforms other surface haptic technologies for its texture rendering capability, as it has a wide bandwidth, generates localized skin vibrations across the touchscreen, and the produced sensations are independent of the screen size~\cite{bau2010teslatouch}. 

Despite its advantages, rendering realistic textures with electrovibration remains challenging. Electrostatic force generation induces a mismatch between the frequency spectra of the input voltage and the resulting friction signals~\cite{vardar2017effect, icsleyen2019tactile}. One reason for this mismatch is the nonlinear behavior of the electrostatic forces in a parallel plate capacitor system, where the generated force is proportional to the square of the input voltage (see Section~\ref{sec:methods} for more details). Additionally, the finger-surface contact acts as an electrical high-pass filter, attenuating low-frequency components in the generated forces~\cite{meyer2013fingertip,vezzoli2014electrovibration, shultz2015surface, vardar2021finger, aliabbasi2022frequency}. Moreover, the mechanical behavior of the fingertip skin, which was shown to behave as a low-pass filter~\cite{hajian1997identification,wiertlewski2012mechanical}, further contributes to this mismatch. 

Achieving accurate texture rendering with electrovibration devices requires calibrating the input voltage to account for the input-output spectral discrepancy. Meyer et al.~\cite{meyer2014dynamics} addressed the electrical high-pass behavior by modulating the amplitude of a high-frequency voltage signal with the desired frequency spectrum. Later, Shultz et al.~\cite{shultz2018application} solved the input squaring by applying the square root operation to the envelope of the amplitude-modulated signal. Subsequently, Grigori et al.~\cite{grigorii2020closed} built a closed-loop controller, dynamically measuring the friction force and controlling input current to maintain a constant voltage, with amplitude modulation, and achieved a theoretical 90\% texture tracking by the controller. However, despite the large variances observed in the experiments, the controller was designed with a single gain term for the input current-to-friction relationship. These variations were attributed to differences in exploration conditions and skin mechanics; failure to account for these could make the controller unstable. Since addressing the input current-to-friction variation was complex, the same authors in the later study~\cite{grigorii2021data} used an open-loop system and constant exploration conditions for texture rendering and achieved 71\% perceptual texture similarity.

Although often overlooked, sliding speed and applied force greatly contribute to the spectrum mismatch for the desired friction on electrovibration devices~\cite{POOL2024FITS}. The soft multilayered structure of the fingertip~\cite{rogelj2022anatomically} influences its dynamic response~\cite{serhat2021free}, with behavior highly dependent on both speed and force~\cite{hajian1997identification, greenspon2020effect}. Previous studies have demonstrated that the frequency composition of skin vibrations shifts toward higher frequencies with increasing speed~\cite{bensmaia2003vibrations,manfredi2014natural}. Similarly, the magnitude of the skin vibrations reduces with an increase in finger normal force~\cite{fagiani2012contact}. Additionally, the skin's mechanical properties -- mass, damping, and stiffness -- rise with the increase in applied force, affecting its vibration spectrum~\cite{wiertlewski2012mechanical}. Applied normal force and sliding speed also cause variations in the air gap, leading to variations in generated electrostatic forces~\cite{guo2019effect,ayyildiz2018contact, vardar2017effect, vardar2021finger}. However, despite these observations, a thorough analysis of how exploratory forces and speeds affect electrostatic forces and how to compensate for these effects in texture rendering has yet to be performed.

In this paper, we investigate how exploration conditions influence the friction between a finger and an electrovibration display, as well as the role of skin mechanics in this interaction, with particular emphasis on their implications for texture rendering. We measured the generated friction forces and skin vibrations when 10 participants explored an electrovibration display under controlled conditions, varying both applied force (0.2~N, 0.3~N, 0.4~N, 0.5~N, and 0.6~N) and sliding speed (20~mm/s, 40~mm/s, 60~mm/s, 80~mm/s, and 100~mm/s). 
The display was driven with 15 logarithmically spaced input frequencies (30~Hz $\sim$ 2~kHz) modulated using the technique described by Shultz et al.~\cite{shultz2018application}. We then identified and modeled the relationship between input voltage and measured friction force, as well as the mechanical response of the finger to the generated friction. For both models, we analyzed how the model parameters varied as a function of applied force and sliding speed. Additionally, since skin biomechanical properties vary among individuals, we examined model variability across participants. Finally, we investigated the relationship between the measured skin mechanical properties and the corresponding model parameters of the electrovibration-induced finger friction response. 



\section{Methodology}\label{sec:methods}

\subsection{Electrovibration}
Electrovibration modulates friction between a capacitive touchscreen and a moving finger using electrostatic forces. The resulting friction force, $F_f$, is influenced by both the applied normal force exerted by the finger, $F_n$, and the electrostatic force, $F_e$. The relation, $F_f=\mu(F_n+F_e)$, describes the friction force, where $\mu$ represents the friction coefficient. The $F_e$ depends on the applied input voltage, $V_i$, and is given by the equation~\cite{shultz2015surface}:
\begin{equation}
    F_e = \frac{A \epsilon_o \epsilon_g }{2}\left(\frac{V_i}{d_g}* \left|\frac{Z_g(\omega)}{Z_g(\omega)+Z_d(\omega)+Z_{sc}(\omega)}\right| \right)^2
    \label{Eqn: Electrovibration3}
\end{equation}
where $A$ is the finger's contact area, $\epsilon_o$ is the permittivity of free space, $\epsilon_g$ and $d_g$ are the permittivity and thickness of the air gap. $Z_g(\omega)$, $Z_d(\omega)$, and $Z_{sc}(\omega)$ represent the electrical impedance of the air gap, the dielectric layer of the touchscreen, and the stratum corneum, respectively.

According to Equation~\eqref{Eqn: Electrovibration3}, the electrovibration force is proportional to the square of the input voltage, expressed as $F_e = kV_i^2$, where $k$ is a constant~\cite{vardar2021finger}. This quadratic relationship introduces nonlinear behavior, causing the output force to oscillate at twice the input frequency. This frequency doubling expands the input spectrum, causing a discrepancy between the input voltage signal and the output skin vibration. See \ref{sec:appA} for derivation.

Another challenge in electrovibration is charge leakage from the conductor to the finger, caused by the electrical properties of the finger-surface contact interface~\cite{vezzoli2014electrovibration,persson2021general,aliabbasi2022frequency}. When a low-frequency voltage input is used as an excitation signal, charges have enough transit time to move from the touchscreen's conductor to the finger through the ridges and vice versa. This reduces the voltage drop across the air gap, reducing the electrovibration force intensity. This leakage significantly attenuates low-frequency components, causing the input voltage to output friction response to exhibit a high-pass filtering effect~\cite{meyer2013fingertip, vardar2017effect}.

As mentioned earlier, eliminating this nonlinearity and any charge leakage is necessary for minimizing the discrepancy between the input voltage and output friction for realistic texture rendering. Due to these physical phenomena, the input voltage spectrum is attenuated at low frequencies and spreads, causing the output friction spectrum to mismatch with the input voltage spectrum. As humans are sensitive to tactile frequencies up to 1~kHz~\cite{bolanowski1988four}, any output friction distortions below 1~kHz may cause a different texture feel than intended, reducing the virtual texture realism. 

One way to mitigate charge leakage and nonlinearity is through amplitude modulation followed by a square root operation on its envelope. In amplitude modulation, the input message signal, $V_m(t)$, is modulated with a high-frequency carrier, $C(t) = \cos(2 \omega_c t )$,~\cite{kang2016investigation}. The bandwidth of the message signal is from 1~Hz to 1~kHz, $\omega_m = 2 \pi f_m \leq 1$~kHz, where $f_m$ and $\omega_m$ are the message signal's frequency and angular frequency. The carrier frequency is typically above 7~kHz so that humans perceive the envelope of the signal and not the carrier itself, $\omega_c = 2 \pi f_c \geq 7$~kHz, where $f_c$ and $\omega_c$ are the carrier signal's frequency and angular frequency. Also, the carrier distorts the message signal if it is too close to the message signal's bandwidth. The amplitude-modulated input signal, $V_i(t)$, has three frequency components -- $\omega_c - \omega_m$, $\omega_c$, and $\omega_c + \omega_m$ -- in the pass-band of the high-pass input voltage to output friction response. During finger electrovibration interaction, the doubling of the input frequency components results in output frequency components at $2\omega_m$ and $2\omega_c$. The finger attenuates any frequency component above 1~kHz~\cite{morioka2005thresholds}, resulting in output friction forces at $2\omega_m$. To ensure the output friction occurs at $\omega_m$, we shift the message by its minimum, $|V_m(t)|$, and apply a square root operation to the envelope of amplitude-modulated signal as shown in Equation~\eqref{eqn: AM signal}. Refer to \ref{sec:appA} for further details and mathematical derivations. 

\begin{figure}[!ht]
    \centering
    \includegraphics[width=\linewidth]{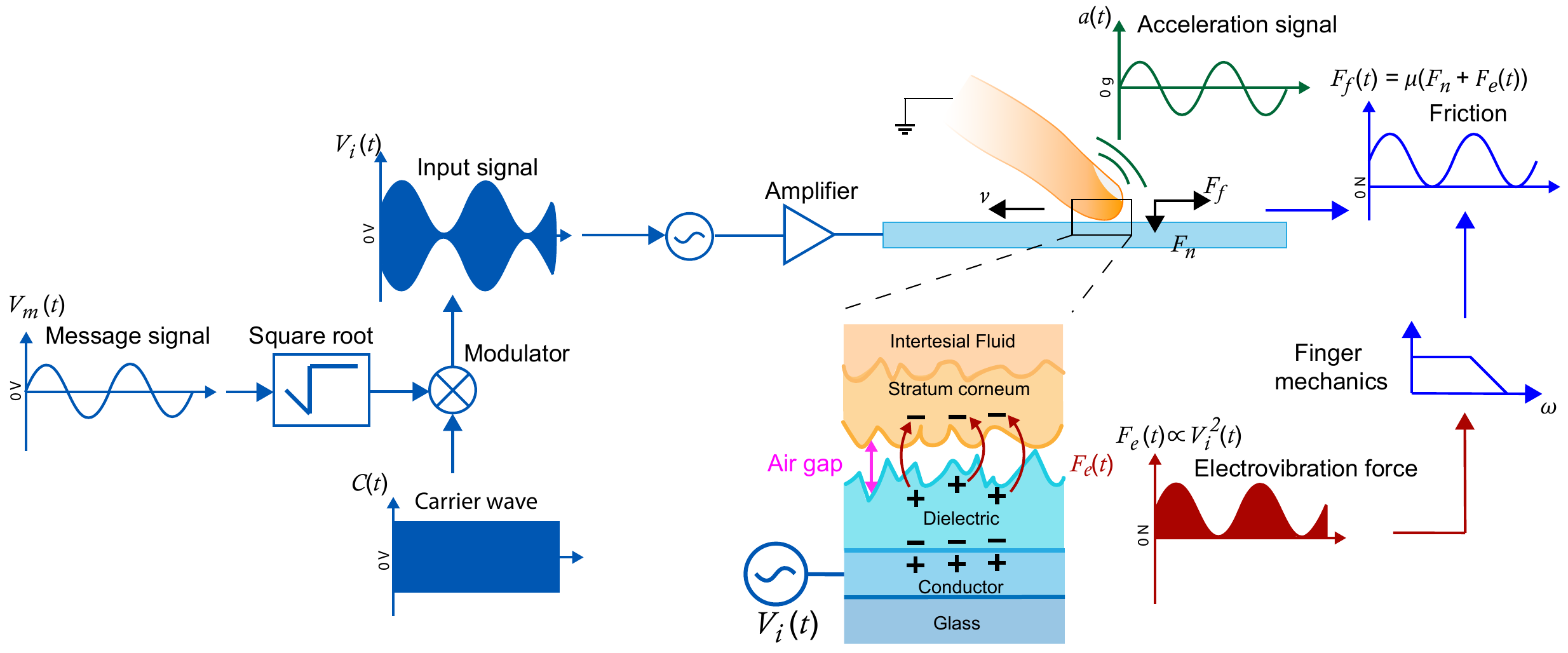}
    \caption{An illustration of how a message signal, $V_m(t)$ is modulated with a carrier signal, $C(t)$, to generate the input voltage, $V_i(t)$. The input signal is applied to a touchscreen that generates electrovibration force, $F_e(t)$, resulting in friction, $F_f(t)$, and skin vibration as an acceleration signal, $a(t)$, during finger sliding.}
    \label{fig: Setup}
\end{figure}

\begin{equation}
    V_i(t) = \sqrt{V_m(t) + |\min(V_m(t))|} \cos(2\omega_c t)
    \label{eqn: AM signal}
\end{equation} 

An illustration describing the process of how an amplitude-modulated input voltage applied to a touchscreen generates electrovibration forces, resulting in friction and skin vibration, is shown in Figure~\ref{fig: Setup}. 

\newpage
\subsection{Experimental setup}

\begin{figure}[!b]
    \centering
    \includegraphics[width=0.8\linewidth]{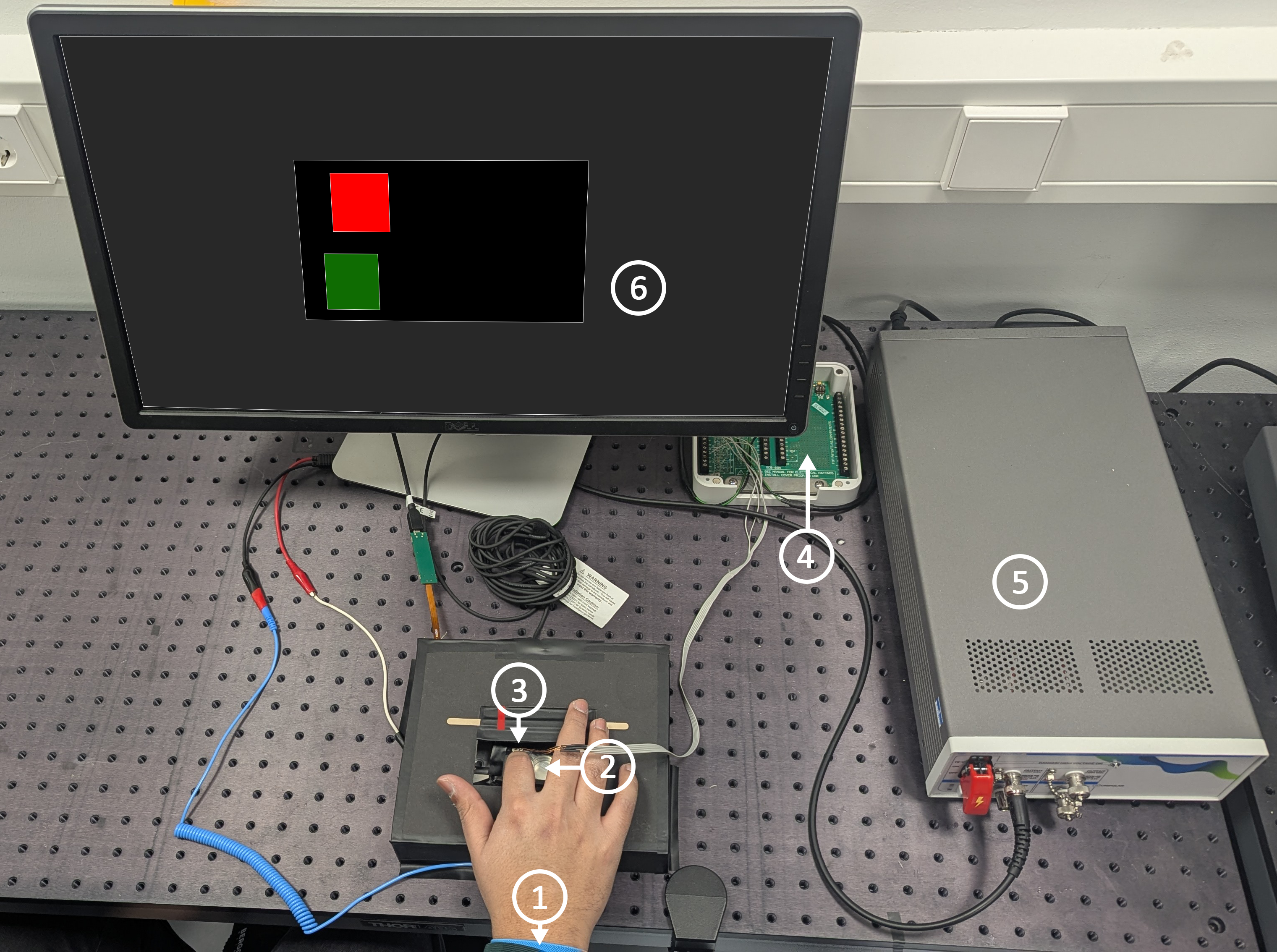}
    \caption{Experimental setup: (1) grounding strap, (2) force sensor and touchscreen, (3) accelerometer, (4) data acquisition card connector, (5) amplifier, (6) monitor displaying the experiment graphical user interface.}
    \label{fig: Setup_actual}
\end{figure}

We used a custom setup (see Figure~\ref{fig: Setup_actual}). The setup featured a touchscreen (SCT3250, 3M), which we cut to a size of $35\times70$~mm. The electrostatic forces were generated by sending voltage signals to the conductive layer of this touchscreen. The voltage signals were designed using a PC (Precision T5820, Dell) and generated by a data acquisition card (PCIe-6323, NI) with a sampling frequency of 20~kHz. These signals were then amplified through a high-voltage amplifier (9200A, Tabor Inc.) with a gain of 50 before sending them to the touchscreen. 

The touchscreen was placed on a force sensor (Nano43 Titanium, ATI) using double-sided tape (Tesa PRO) to measure the generated forces in three dimensions. We mounted the force sensor on a large aluminum block with screws. This entire assembly was placed on an aluminum optical breadboard table (MB60120/M, Thorlabs), supported by a sturdy frame (PFM52502, Thorlabs). A monitor (UltraSharp 24, Dell) displayed graphical information to the participants. The data acquisition card collected data from the force sensor at a sampling rate of 20~kHz. A 2D infrared position sensor (NNAMC1580PCEV, Neonode) tracked the finger's position with a resolution of 0.1~mm and a sampling rate of 60~Hz. The sensor was mounted on a 3D-printed support, ensuring its active plane was positioned above the touchscreen. We enclosed the touchscreen in a box with a 70$\times$80~mm cutout for finger exploration. We also measured the three-dimensional skin vibrations via an accelerometer (ADXL358, Analog Devices) attached to the participants' fingernails via an accelerometer wax (Model 080A109, PCB Piezotronics). The accelerometer data was also collected through the same data acquisition card with a sampling rate of 20~kHz.

We designed the experimental setup to ensure the mechanical resonance frequency outside the tactile range (1~Hz to 1~kHz). As the experimental setup has peak sensitivity at its resonance frequency, any input force at this frequency would be measured as higher than the original signal intensity. 
Hence, a resonance frequency below 1~kHz would lead to an error in the measured force signals in the tactile range. To verify the setup's bandwidth, we struck the touchscreen with an impact hammer (086E80, PCB Piezotronics) along the normal and lateral directions and measured the response using the force sensor beneath the screen. The results indicated a resonance frequency of 1.4~kHz in the normal direction (see Figure~\ref{fig: Normal_response}) and two resonances at 866~Hz and 1740~Hz in the lateral direction (see Figure~\ref{fig: Lateral_response}).

\subsection{Procedure}
We conducted the experiments following the ethical principles of the Declaration of Helsinki. We obtained approval from the TU Delft Human Research Ethics Committee (application number 3469). The study involved five women and twelve men, with an average age of 26.3 years (SD: $\pm$2.36 years). All participants provided informed consent before the experiments. 

The aim of the experiments was to record the finger interaction forces and accelerations while the participants explored the electrostatically actuated touchscreen at a 60$^{\circ}$ contact angle. This specific angle was chosen based on prior research indicating that a 60$^{\circ}$ contact angle naturally occurs during texture discrimination tasks~\cite{smith2002deployment,callier2015kinematics}. Furthermore, higher contact angles have been shown to increase finger stiffness, thereby enhancing the transmission of vibratory cues to the fingertip~\cite{babu2018introducing,louyot2024influence}. This improvement in vibration transmission contributes to better texture perception, which is relevant for texture rendering. The cut-out of the enclosure box around the touchscreen ensured that the finger always maintained a constant contact angle with the touchscreen (see Figure~\ref{fig: Setup_actual}). 

During the experiments, a participant moved the index finger of their dominant hand on the touchscreen horizontally (left to right) at five different sliding speeds (20~mm/s, 40~mm/s, 60~mm/s, 80~mm/s, and 100~mm/s) while applying five different normal forces (0.2~N, 0.3~N, 0.4~N, 0.5~N, and 0.6~N). These force and speed values were selected based on previous studies that reported typical ranges observed during natural active touch exploration -- approximately 0.15~N to 0.64~N for normal force~\cite{smith2002deployment, louyot2024influence,kejriwal2023user}, and 10~mm/s to 100~mm/s for sliding speed~\cite{callier2015kinematics, greenspon2020effect,kejriwal2023user}. Horizontal movement was specifically chosen because the finger predominantly exhibits rolling motion in this direction, which is associated with a lower coefficient of friction and reduces the likelihood of stick-slip behavior~\cite{nakanishi2025friction}. 

A custom graphical user interface (GUI) guided the participants by displaying two visual square cues: the top square indicated the reference speed, while the bottom square tracked the user's finger movement (Figure~\ref{fig: Setup_actual}). The tracking square changed color to indicate the applied force: green for the target range, yellow for below the target, and red for above the target. If the participant's mean, minimum, and maximum exploration force and speed deviated by $\pm10\%$ and $\pm25\%$ of the target range for more than one second, then the trial was discarded and repeated. More details regarding this procedure can be found in \ref{sec:AppC}.

Each participant explored the touchscreen at different speeds and forces while it was excited with 15 logarithmically spaced sinusoidal signals ranging from 30~Hz to 2~kHz. Each signal was amplitude-modulated with a 7~kHz carrier frequency, as defined in Equation~\eqref{eqn: AM signal}. The peak-to-peak amplitude of the final input voltage signal was 150~$V_{pp}$.

Each trial lasted 17 seconds and consisted of two phases. During the initial 7 seconds, the participant adjusted their finger movements to match the experimental conditions; however, no interaction data was recorded during this period. In the subsequent 10 seconds, we measured the friction between the fingertip and the touchscreen using a force sensor mounted beneath the surface, as participants slid their finger at the instructed speed and applied normal force. This measurement captured both the baseline friction with the smooth glass and the additional friction induced by electrovibration. Concurrently, we recorded fingertip skin vibrations using an accelerometer attached to the participant's fingernail. Participants typically completed more than six swipes in 10 seconds—--approximately three lateral and three medial. At the lowest speed (20~mm/s), each 70~mm swipe took around 2.35~s, reducing the average to three lateral and two medial swipes. The procedure was repeated for each subsequent trial.

This experimental configuration resulted in 375 trials (15 frequencies $\times$ 5 speeds $\times$ 5 forces) for each participant. After every 15 trials, we gave the participants a 10-second break. Each participant spent approximately 6 hours completing data collection. Due to the long experiment duration, the participants performed the experiment in multiple sessions.

Some participants did not have good hand-eye coordination and failed to meet the target range for the experimental conditions. We rejected the data from seven participants who spent more than one hour on a trial set of one sliding speed and five exploration forces. In the end, three women and seven men successfully completed the experiment, whose data was further processed.

\begin{figure}[!b]
    \centering
    \includegraphics[width=\linewidth]{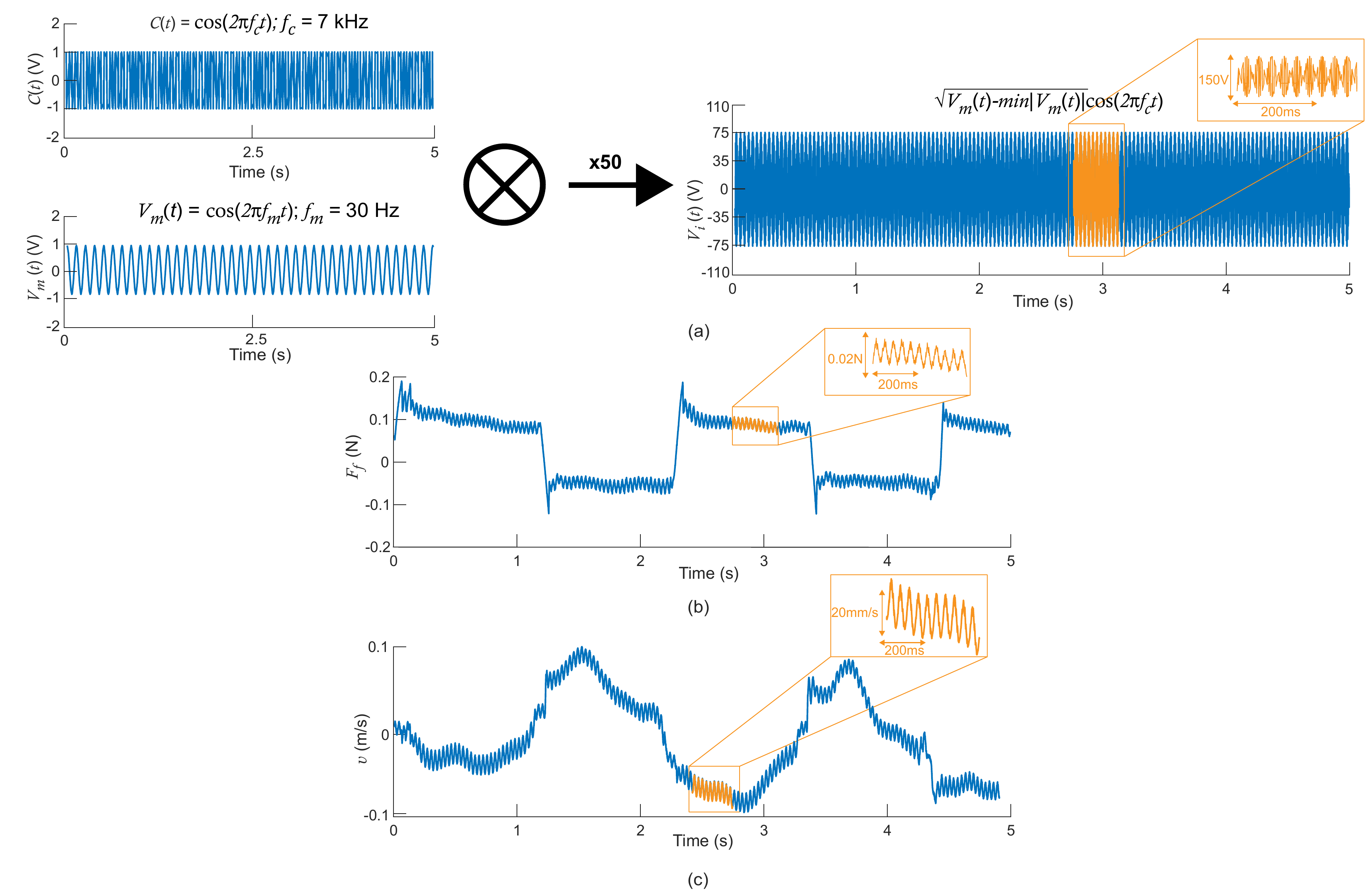}
    \caption{An example of experimental data collected during the finger-touchscreen interaction at a speed of 40~mm/s and a normal force of 0.2~N. A 30~Hz message signal was shifted with its minimum and modulated with a 7~kHz carrier signal. The resulting amplitude-modulated signal's envelope was square-rooted and then amplified to 150~V (peak to peak) to excite the touchscreen. The plots show raw data, with the highlighted sections extracted for analysis. The recorded (a) amplitude-modulated input voltage signal, (b) 1D friction signal obtained through dimension reduction of measured lateral forces, and (c) finger vibration as the integrated velocity signal derived from the raw acceleration data.}
    \label{fig: Raw data}
\end{figure}

\subsection{Data processing}

During the experiments, we recorded the contact forces, finger accelerations, and the input voltage signal after modulation for analysis, as shown in Figure~\ref{fig: Raw data}(a). The finger's active exploration generally did not follow a perfectly straight path (see Figure~\ref{fig: Finger_trajectory}). The resulting lateral forces were primarily captured along the force sensor's y-axis, with partial contributions along the x-axis. To capture the generated friction forces accurately, we transformed the two-dimensional lateral force data into a single dimension in the frequency domain~\cite{landin2010dimensional}, extracting the friction signal as shown in Figure~\ref{fig: Raw data}(b). Similarly, the accelerometer captured finger vibrations caused by friction along its three axes, with the x-axis primarily recording lateral vibrations in the direction of finger exploration and residuals along the y and z axes. To align the accelerometer axes with the force sensor axes, we multiplied two rotation matrices: one along the x-axis at $155^{\circ}$ and another along the y-axis at $180^{\circ}$, accounting for the inverse orientation of the accelerometer on the y-axis and the finger exploration angle of $60^{\circ}$. After converting raw sensor readings to true acceleration values by multiplying them by $9.8 m/s^2$, we applied the DFT321 algorithm~\cite{landin2010dimensional} to reduce the 3D signals and obtain the resultant acceleration along the exploration direction. This signal was then integrated over time to obtain the velocity data, as shown in Figure~\ref{fig: Raw data}(c). 

From the input voltage signal (amplitude modulated), the friction force, and the finger vibration signal, we extracted sections to identify mathematical models. We used the friction data as a reference and set a threshold condition to determine when the participants made a finger sweep from left to right. If the friction force was above its mean value, then the participant would sweep from left to right. Next, we found the start and end indices of all the left-to-right sweeps made by the participants. We then averaged a particular sweep's start and end index to find its middle index. We proceeded to extract samples of 1999 long to the left and 2000 long to the right of the middle index. By using the time stamps of the 4000 samples long data, we extracted the corresponding data sections from the amplitude-modulated signal and the finger vibration signal as shown by the highlighted sections in Figure~\ref{fig: Raw data}. We selected the sample size of 4000 because it is a multiple of the sampling rate 20~kHz, which ensures all test sinusoidal frequencies appear in the Fourier transform. Additionally, at the maximum sliding speed of 100~mm/s, the recorded data was 7500 samples long, so we used 4000 samples for consistency across all exploration conditions.

After extracting the samples for the input amplitude-modulated voltage signals, $V_i$, for different finger sweeps, we demodulated them to get the original message signals back to model the input message signal, $V_m$, to output the friction, $F_f$, response. To do this, we applied the Fourier transform on the input amplitude-modulated signal, $V_i$, and then we calculated the magnitude and phase of the upper envelope, $\omega_c+\omega_m$, of the amplitude-modulated signal. Using the magnitude and phase value, we reconstructed the sinusoidal message signal. We repeated the same process for all the sweep data, frequencies, exploration conditions, and participants. 

\subsection{Dynamic modeling of electrovibration-induced finger friction on a touchscreen}
\begin{figure}
    \centering
    \includegraphics[width=0.85\linewidth]{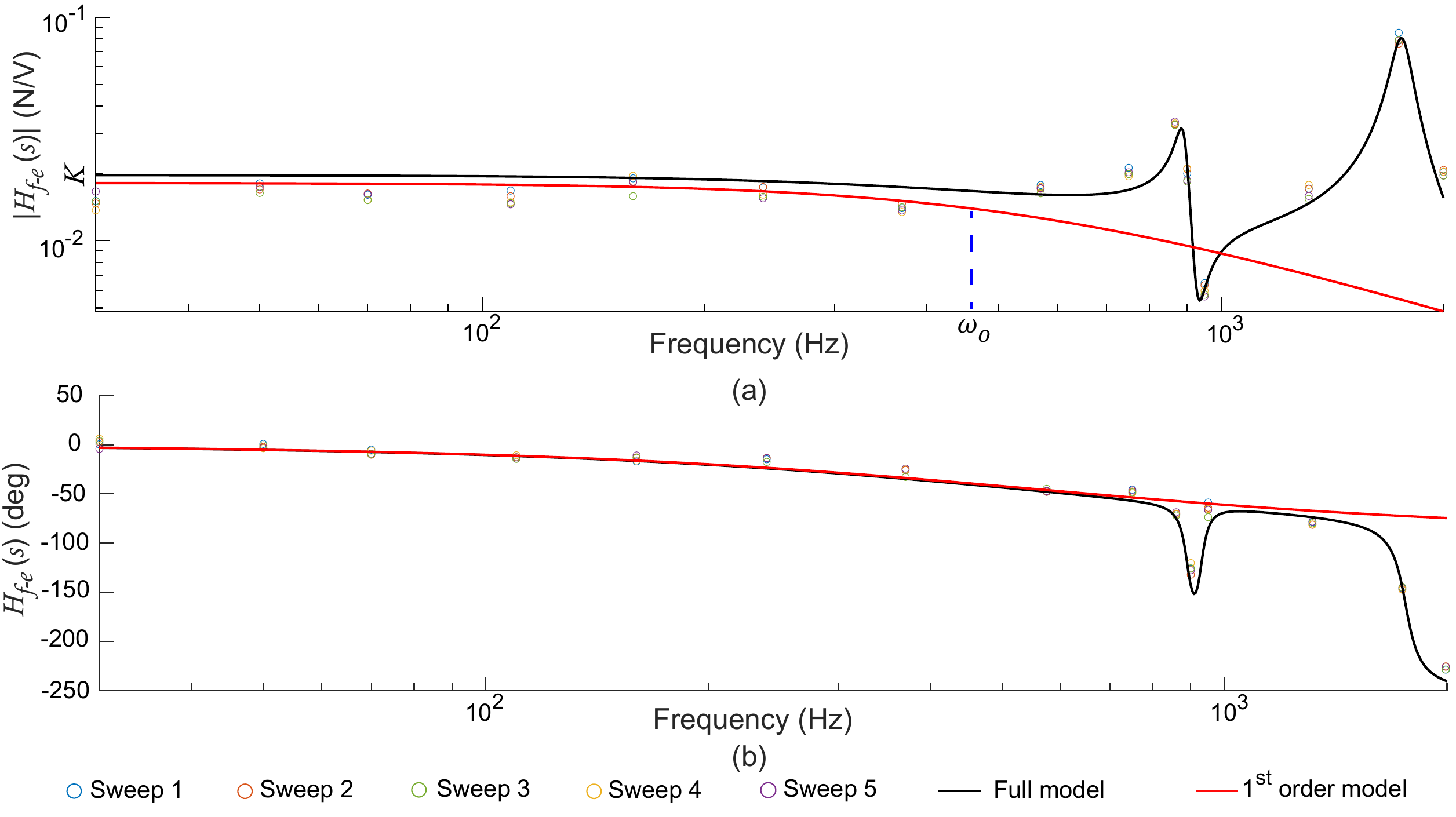}
    \caption{Bode plot of the electrovibration-induced finger friction on a touchscreen, $H_{f-e}$, for input voltage envelope when the display was explored with a sliding speed of 40~mm/s and an applied force of 0.2~N. Each point in the figure represents the Fourier transform when the output friction force, $F_f$, is divided by the input message signal, $V_m$, at that frequency. Each color shows the sweep order made by the finger at a given trial. The response up to 750~Hz reflects the electrovibration-induced finger friction, while the mechanical response of the measurement setup dominates frequencies beyond 750~Hz. The black curve represents the model incorporating both the finger and setup response. The red curve represents the first-order electrovibration-induced finger friction response after removing the setup's behavior.}
    \label{fig: Model data}
\end{figure}

Our goal was to investigate how exploration speed and force influence the electrovibration-induced finger friction on a touchscreen, with the primary aim of understanding this behavior in the context of texture rendering. To achieve this, we modeled the relationship between the input message signal, $V_m$, and output friction, $F_f$, while assessing the impact of exploration conditions on the model parameters. Given the nonlinear behavior of the fingertip when interacting with a surface~\cite{miguel2015characterization}, we sought to simplify the electrovibration-induced finger friction model by conditionally representing it as a linear system. To check this, we identified the transfer function, $H_{f-e}(s)$, by dividing the Fourier transform of the output friction, $F_f$, by the Fourier transform of the input message signal, $V_m$, for a particular exploration force and speed. We repeated this process for every finger sweep made by the participant and every test frequency to construct a Bode plot as shown in Figure~\ref{fig: Model data}; the input-output dynamics showed a 20~dB/ decade roll off rate matching a first-order low-pass up to 750~Hz and then showed two resonance peaks at 860~Hz and 1740~Hz. We verified that the two resonance peaks are due to the experimental setup's hardware; please refer to \ref{sec:AppB} for more details. We used the fmincon algorithm in MATLAB to optimize the transfer equation based on the discrete transfer function points found earlier. Then, we removed the setup's dynamics as given by Equation~\eqref{eqn: setup response} from the fitted transfer function model. As a result, we found a first-order response for the input message signal, $V_m$, to output friction, $F_f$, response, as shown in Equation~\eqref{eqn: EV response}.
\begin{equation}
    H_{f-e}(s) = \frac{F_f(s)}{V_m(s)} = \frac{K\omega_o}{s+\omega_o}
    \label{eqn: EV response}
\end{equation}

Here, $F_f(s)$ and $V_m(s)$ are the output friction and input message signal in the frequency domain, $K$ is the system's gain, and $\omega_o$ is the cutoff frequency. 

This first-order behavior aligns with previous findings by Meyer et al.~\cite{meyer2014dynamics}, who demonstrated that the message signal of an amplitude-modulated voltage signal and the resulting friction force, $F_f$, exhibited a first-order low-pass response at an arbitrary exploratory force and speed. They modeled the behavior using the device's quality factor, $Q$, and carrier frequency, $\omega_c$, as $H_{f-e} = \frac{Q}{1+{2Qs}/{\omega_c}}$. However, to better represent the input message signal, $V_m$, to output friction, $F_f$, response variation caused by exploration speed and force, we opted for a generic model given by Equation~\eqref{eqn: EV response}. Figure~\ref{fig: Model data} shows an input message signal, $V_m$, to output friction, $F_f$, response computed for one of our participants at 40~mm/s exploration speed and 0.2~N normal force. The corresponding estimated model parameters are $K = 0.0147$~(N/V), and $\omega_o = 961$~Hz.

\begin{figure}[!ht]
    \centering
    \includegraphics[width=\linewidth]{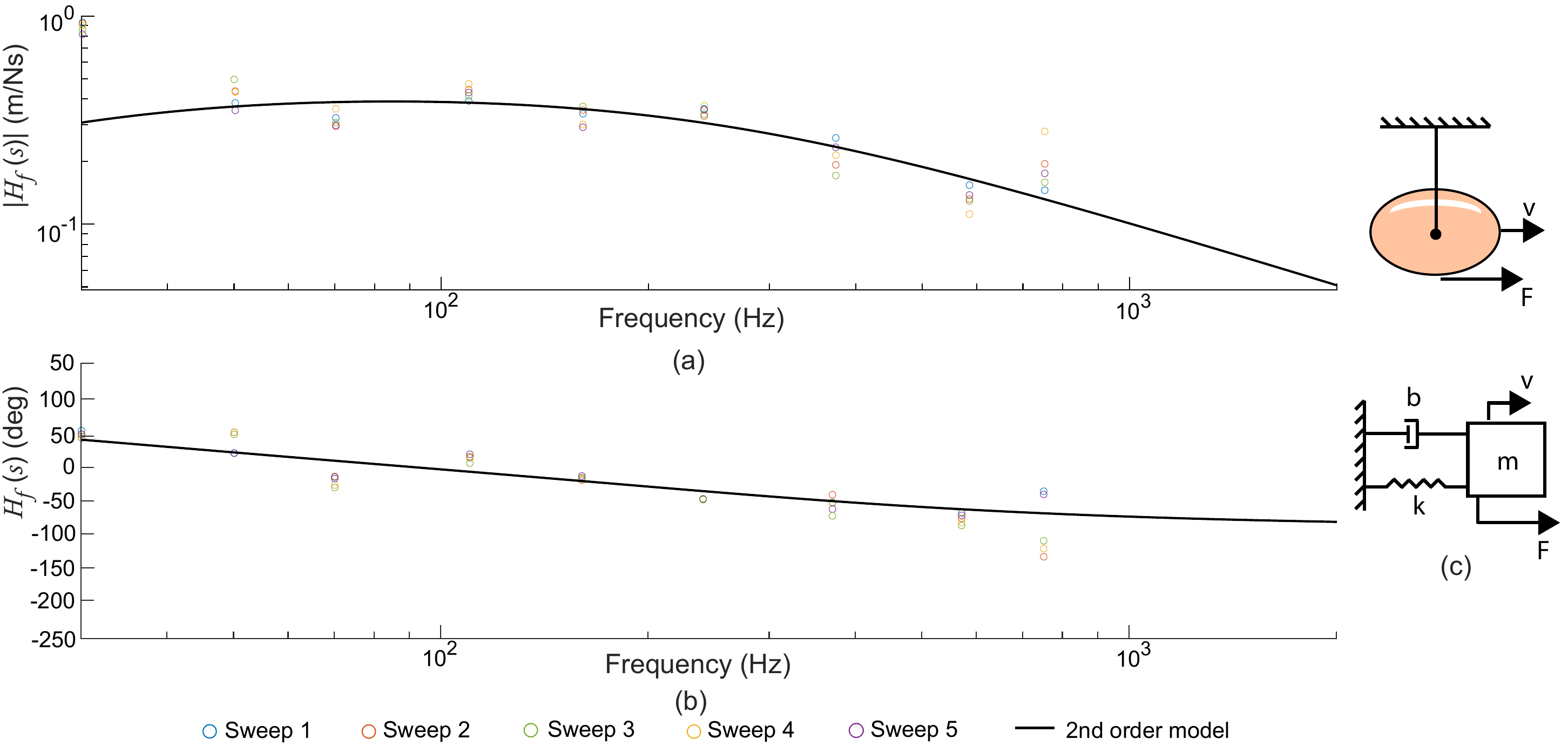}
    \caption{Bode plot of the finger mechanical response, $H_{f}$, for input electrovibration-induced finger friction when the display was explored with a sliding speed of 40~mm/s and an applied force of 0.2~N. Each point in the figure represents the Fourier transform when the output finger vibration velocity is divided by the input finger friction force. Each color shows the sweep order made by the finger at a given trial. The model analysis is limited to 750~Hz due to the resonance of the setup.}
    \label{fig: Mechanical model}
\end{figure}

While higher-order models can also be used to capture the more complex dynamics of electrovibration-induced finger friction, we opted for a low-order model. Higher-order models risk overcompensation during texture rendering, potentially leading to instability and unnecessary complexity. In contrast, low-order models offer a more robust and practical solution for real-time rendering applications.

\subsection{Dynamic modeling of finger skin mechanical behavior}
Additionally, we modeled finger vibrations caused by the friction signal to investigate how the finger's mechanical behavior influences electrovibration-induced finger friction on a touchscreen. Here, we considered the measured electrovibration-induced finger friction force, $F_f$, as input and finger vibration velocity, $v$, as output. As the finger exhibits a nonlinear behavior~\cite{bischoff2000finite, miguel2015characterization, d2017preliminary}, we wanted to know if the input friction force, $F_f$, to output finger vibration velocity, $v$, can be represented by a linear system. With this aim, we divided the Fourier transform of the finger vibration signal by the Fourier transform of the friction signal to obtain the transfer function, $H_f$. We repeated the process for every finger sweep and frequency for a particular exploration condition, similar to electrovibration-induced finger friction modeling. After plotting the transfer function points using a Bode plot, see Figure~\ref{fig: Mechanical model}, we observed that the system showed a second-order response. 
Then, using the fmincon algorithm in Matlab, we modeled the system based on the commonly used mass, spring, and damper system in the literature~\cite{hajian1997identification,wiertlewski2012mechanical,grigorii2021data}:
\begin{equation}
    H_f(s) = \frac{v(s)}{F_f(s)} = \frac{s}{ms^2+bs+k}
\end{equation}
Here, $v(s)$ and $F_f(s)$ are the Fourier transform of the finger vibration as velocity signal and friction generated due to finger electrovibration interaction, $H_f(s)$ is the transfer function between them, $m$ is finger-moving mass, $b$ is finger damping, and $k$ is finger stiffness. The moving mass of the finger represents the effective mass of the subcutaneous finger pad, while the damping and stiffness correspond to the viscosity and elasticity of the fingertip pulp~\cite{hajian1997identification}. Figure~\ref{fig: Mechanical model} shows an example finger mechanical response obtained when the display was explored with 40~mm/s sliding speed and 0.2~N applied force. The finger mechanical model parameters for the response are $m = 0.0015$~Kg, $b = 1.3$~Ns/m, $k = 444$~N/m. Our findings matched the mechanical response, their derived mass, spring, and damping values reported in the previous studies~\cite{hajian1997identification,wiertlewski2012mechanical,grigorii2021data}, despite differences in their excitation and measurement techniques.

\section{Results}\label{sec:results}

\subsection{Dynamics of electrovibration-induced finger friction on touchscreens}

\begin{figure}[!h]
    \centering
    \includegraphics[width=\linewidth]{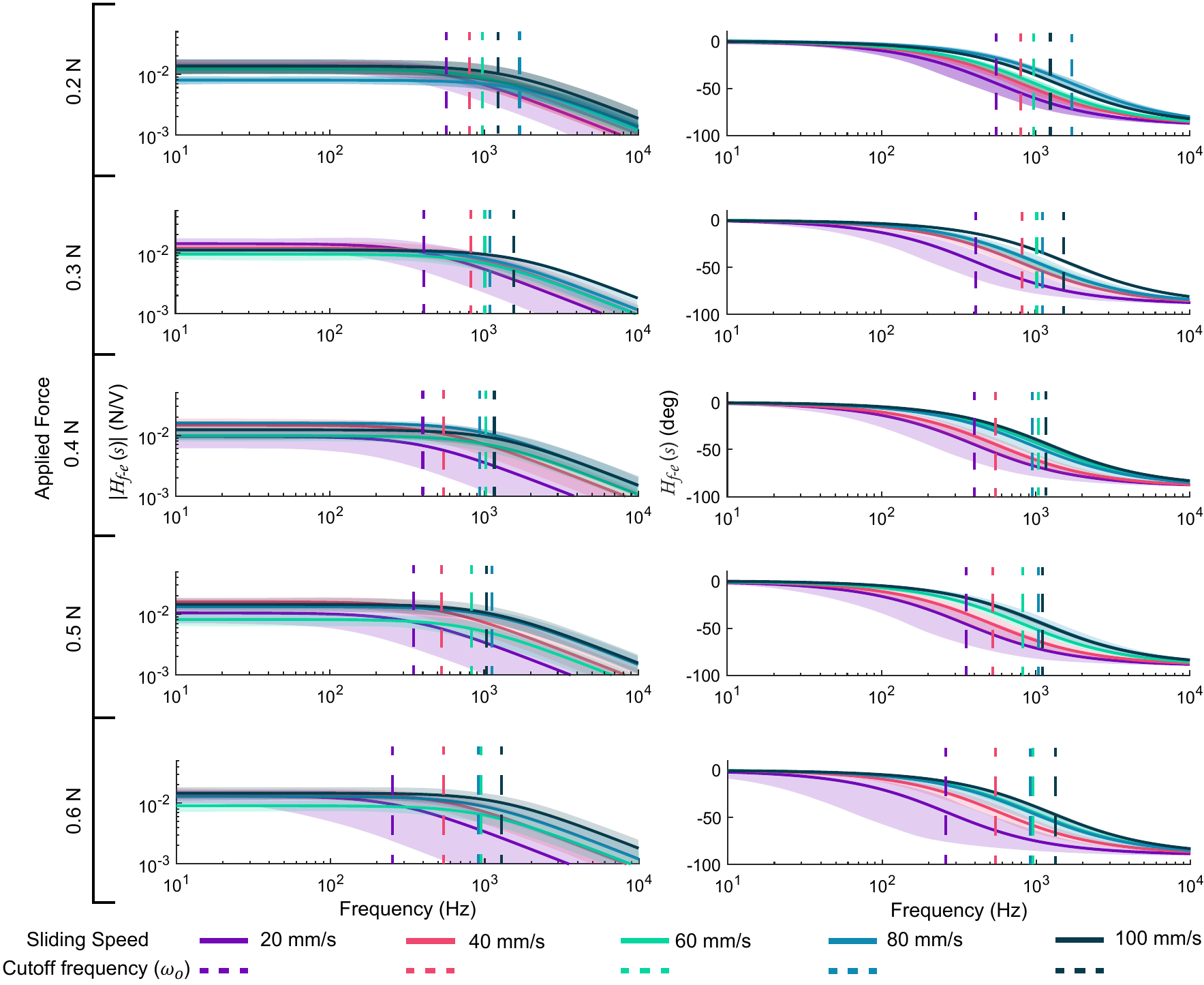}
    \caption{The obtained transfer functions, $H_{f-e}$, representing the relationship between the measured friction force, $F_f$, and the input message signal, $V_m$, across tested exploration speeds and forces for all participants. The color-coded center lines show the mean transfer function across different speeds, while the shaded region represents the standard error across participants. The dashed vertical lines mark the 3-dB cutoff frequencies, $\omega_o$.}
    \label{fig: elecmech}
\end{figure}

\begin{figure}[!h]
    \centering
    \begin{subfigure}{\textwidth}
        \includegraphics[width=\linewidth]{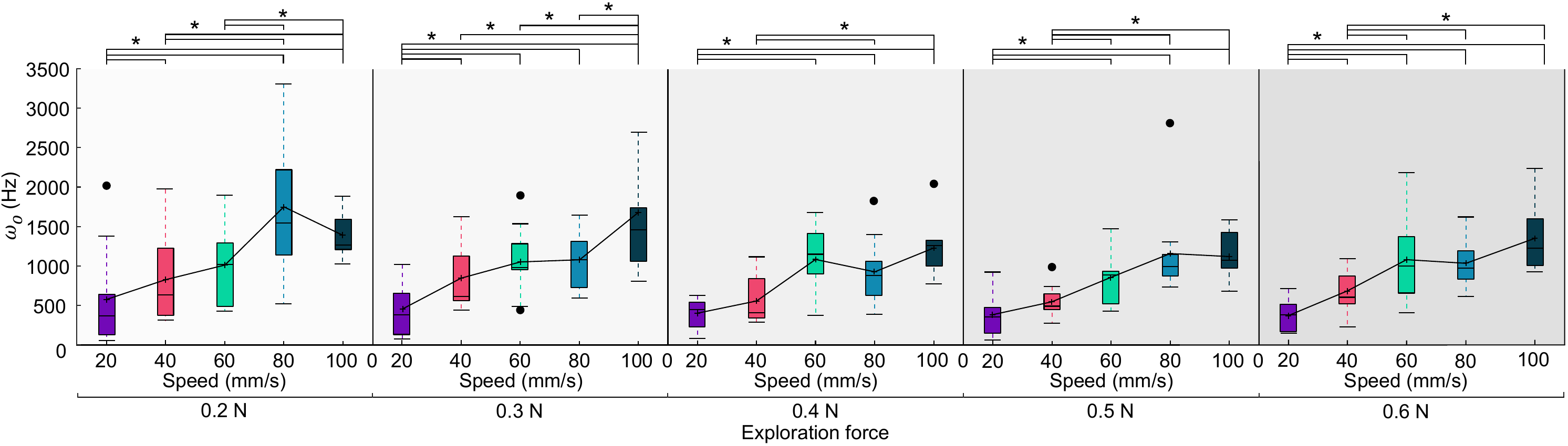}
        \caption{}
        \label{fig: elecmech_gain}
    \end{subfigure}
    \hfill
    \centering
    \begin{subfigure}{\textwidth}
        \includegraphics[width=\linewidth]{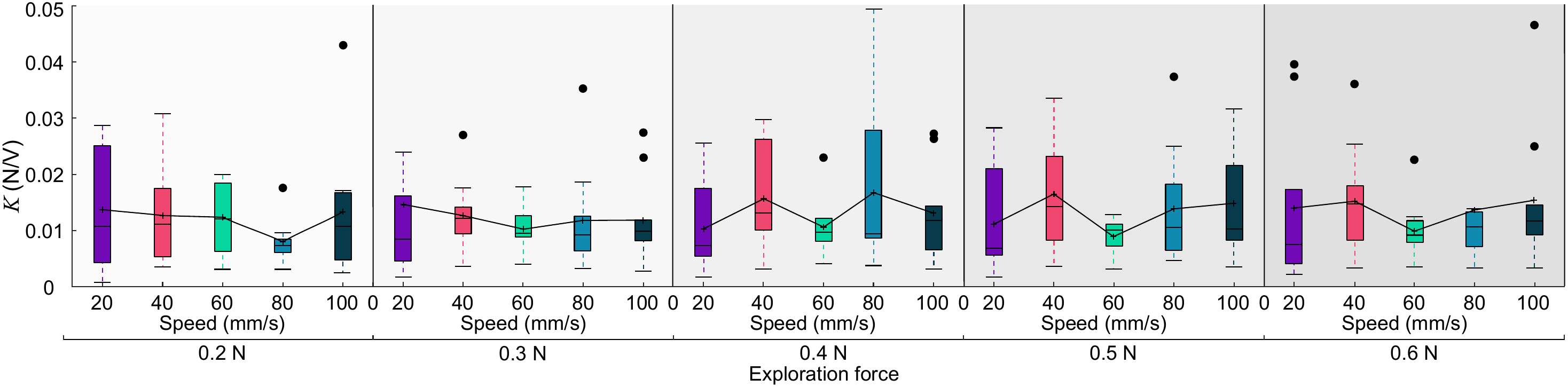}
        \caption{}
        \label{fig: elecmech_cutoff}
    \end{subfigure}
    \caption{The first-order model parameters of the transfer function, $H_{f-e}$, between the measured friction and input message signal, $V_m$, for the tested exploration speeds and forces across participants: (a) gain, $K$, and (b) cutoff frequency, $\omega_o$. The results corresponding to each exploration force and speed are shown in separate color-coded box plots. The center lines show the medians; box limits indicate the 25th and 75th percentiles. The whiskers extend to 1.5 times the interquartile range. The filled black circle ($\bullet$) and the plus (+) symbols indicate the outliers and means, respectively. The asterisk (*) and double asterisk (**) with the connected brackets show the statistically significant pairs at ($p<0.05$) and ($p<0.01$), respectively.}
\end{figure}

The estimated first-order transfer functions, which represent the relationship between the measured friction force, $F_f$, and the input message signal, $V_m$, across different applied forces and sliding speeds, are shown in Figure~\ref{fig: elecmech}. The corresponding estimated model gains and cutoff frequencies are shown in Figures~\ref{fig: elecmech_gain} and~\ref{fig: elecmech_cutoff}. 

We performed statistical analyses to examine how sliding speed and force influence electrovibration-induced finger friction. In our analysis, the applied force, $F_n$, and sliding speed, $\nu$, were treated as independent variables (main effects), while the model parameters -- the gain $K$ and cutoff frequency $\omega_o$ -- were the dependent variables. We assessed the normality of all data using the Kolmogorov-Smirnov test, which indicated that both the gains and cutoff frequencies were not normally distributed ($p < 0.01$). Hence, we applied a generalized linear mixed model (GLMM) to test the influence of the independent variables. To account for the variability in these parameters introduced by different participants, the participant ID number was included as a random effect.
The resulting mixed model with a linear approximation for the gain parameter $K$ is given as:
\begin{equation}
     K = \alpha_o+\alpha_1 F_n+\alpha_2 \nu+\alpha_3F_n*\nu+\rho+ \epsilon
\end{equation}
Here, $\alpha_o$ is the model's intercept parameter, $\alpha_1$, $\alpha_2$, and $\alpha_3$ are the fixed coefficients corresponding to the effects of force, speed, and the force-speed interaction, $\rho$ is the coefficient for the random between-participant effect, and $\epsilon$ is the model's residual. The fixed coefficients show the estimated influence of the independent variable on the dependent variable, with other independent variables being constant. Our analysis revealed that the main effects of applied force ($F(1,178) = 0.310$, $ p = 0.578$) and sliding speed ($F (1,214) = 0.715$, $p = 0.399$), and their interaction ($F(1,210) = 0.892$, $p = 0.346$) all did not affect the gain. 

To evaluate the variability across participants, we conducted a likelihood ratio test ($LRT$)~\cite{pinheiro2000mixed} by comparing the full GLMM model, which included the random effect, with a reduced model excluding it. This procedure involved calculating the -2 log-likelihood values for both the full ($-2LL_{full}$) and reduced ($-2LL_{reduced}$) models, defined as:
\begin{equation}
    LRT = -2LL_{reduced}-(-2LL_{full})
\end{equation}
If the resulting $LRT$ value exceeds the critical value from the chi-square distribution (with degrees of freedom equal to the number of random effects, here $df = 1$), the random effect is considered statistically significant. In our analysis, $-2LL_{reduced} = -1594.327$ and $-2LL_{full}= -1631.082$, yielding an $LRT$ value of 36.755. This number exceeds the critical chi-square value of 3.84 at $p = 0.05$, indicating that the random effect due to participants had a statistically significant impact on the model.

Similarly, the mixed model with linear approximation used for statistical analysis of the cutoff frequency $\omega_o$ data is given as:
\begin{equation}
     \omega_o = \beta_o+\beta_1 F_n+\beta_2 \nu+\beta_3 F_n*\nu+\rho+ \epsilon
\end{equation}

Here, $\beta_o$ is the model intercept, $\beta_1$, $\beta_2$, and $\beta_3$ are the fixed coefficients for force, speed, and the force-speed interaction, $\rho$ is the coefficient for the random effect due to participant ID, and $\epsilon$ is the model's residual. Unlike the statistical analysis results for $K$, the cutoff frequency was significantly affected by the sliding speed ($F(1,211) = 25.068$, $p < 0.001$). Moreover, the random effect due to participants was also statistically significant ($-2LL_{reduced} = 3732.094
$ and $-2LL_{full}= 3698.635$, yielding an $LRT$ = 33.459 $>$ 3.84). However, no significant effects of the applied force ($F(1,176) = 1.12$, $p = 0.291$), and the speed-force interaction ($F(1,207) = 0.715$, $p = 0.399$) were found. 

These analyses show that only the finger speed significantly affects the cutoff frequency $\omega_o$ with a corresponding fixed coefficient, $\beta_2$, value of 13.811~Hz/mm/s. This result indicates that for every 1~mm/s increase in speed, the cutoff frequency increases by 13.811~Hz, relative to an intercept, $\beta_o$, of 385.68~Hz. Multiplying this coefficient by the difference between the lowest and highest speeds yields a total increase of 1104~Hz. Additionally, dividing the coefficient by the mean cutoff frequency at the lowest speed indicates a 2.47\% increase in cutoff frequency for each unit increase in speed. 

Next to the GLMM analysis, we also performed a post-hoc Wilcoxon signed-rank test to evaluate if the cutoff frequencies across the participants statistically differ with sliding speeds for different forces; see bars at the top of Figure~\ref{fig: elecmech_cutoff} for the significantly different pairs. The post-hoc analysis showed a significant difference in cutoff frequency across sliding speeds, particularly between the lower and higher speed ranges. However, as the applied force increased, the cutoff frequency remained insignificant at higher force levels (60–100~mm/s).

Based on our GLMM analysis results, we propose the following empirical model that can be used to predict the electrovibration-induced finger friction dynamics as a function of the sliding speed, $\nu$: 
\begin{equation}
    H_{f-e}(s,\nu) = \frac{K \omega_o(\nu)}{s+\omega_o(\nu)}  \in K = 0.0123; \omega_o =  385.68+13.811\nu
    \label{eqn: finger_display_model}
\end{equation}
Here, $K = 0.0123$ represents the average gain across all sliding speeds and applied forces measured in the experiment, as our GLMM analysis showed no significant effects of these factors on the gain. However, as the cutoff frequency was influenced by the sliding speed, we simplified the corresponding model as $\omega_o(\nu) =  385.68+13.811\nu$, capturing the variation in finger-display interaction with sliding speed. Note that this model is only valid for an input voltage amplitude of 150~$V_{pp}$; it does not capture variations with amplitude changes. Additionally, it reflects an average response, omitting inter-participant variability.     

\begin{figure}[!h]
    \centering
    \begin{subfigure}{\textwidth}
        \includegraphics[width=0.98\linewidth]{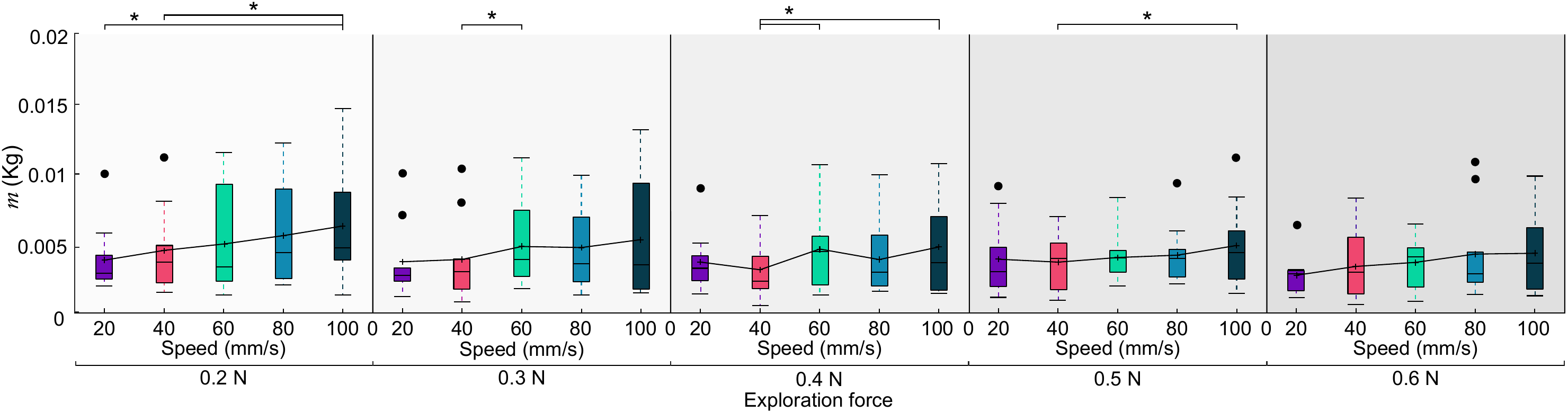}
        \caption{}
        \label{fig: mech_mass}
    \end{subfigure}
    \hfill
    \begin{subfigure}{\textwidth}
        \includegraphics[width=0.98\linewidth]{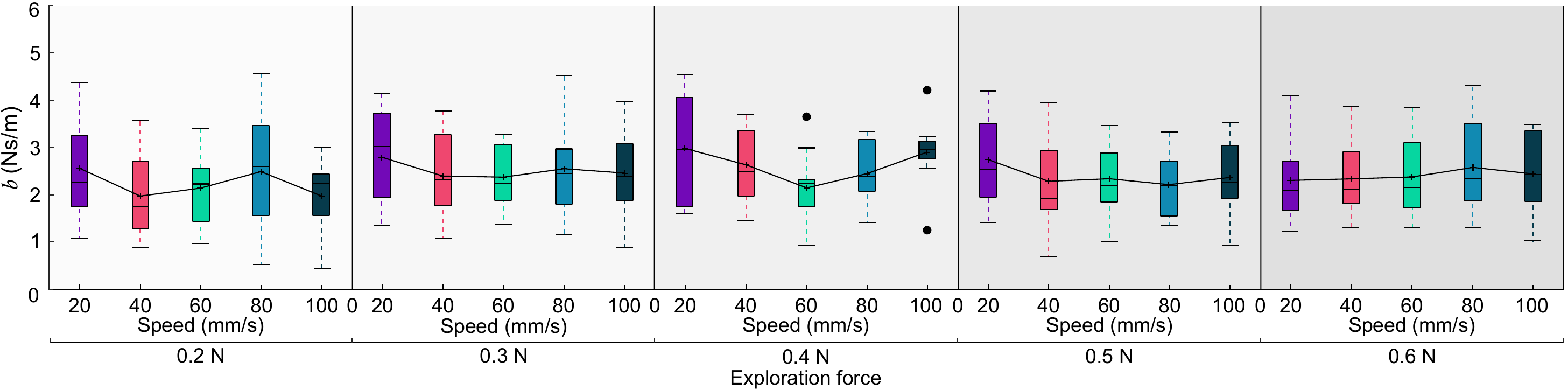}
        \caption{}
        \label{fig: mech_damp}
    \end{subfigure}
    \hfill
    \begin{subfigure}{\textwidth}
        \includegraphics[width=0.98\linewidth]{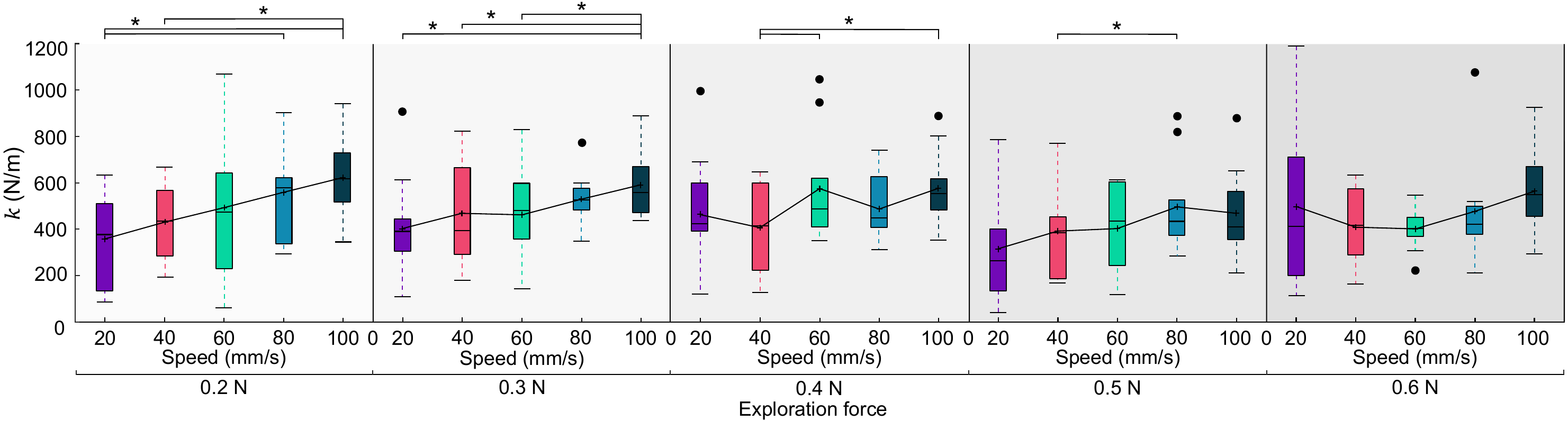}
        \caption{}
        \label{fig: mech_stiff}
    \end{subfigure}
    \caption{Results showing the variation of finger's mechanical response, $H_{f}$, with sliding speed and applied force across participants. Box plots of finger's (a) mass, $m$, (b) damping, $b$, and (c) stiffness, $k$. The results corresponding to each applied force and sliding speed are shown in separate color-coded box plots. The center lines show the medians; box limits indicate the 25th and 75th percentiles. The whiskers extend to 1.5 times the interquartile range. The filled black circle ($\bullet$) and the plus (+) symbols indicate the outliers and means, respectively. The asterisk (*) with the connected brackets shows the statistically significant pairs at ($p<0.05$).}
    
\end{figure}

\subsection{Dynamics of finger's mechanical behavior}

The mass, damping, and stiffness of the second-order model---fitted to the measured finger vibration velocity driven by the electrovibration-induced friction input---are presented in Figures~\ref{fig: mech_mass},~\ref{fig: mech_damp}, and~\ref{fig: mech_stiff}, corresponding to sliding speed and applied force conditions. A Kolmogorov-Smirnov test confirmed that these parameters were not normally distributed ($p<0.01$). Accordingly, we assigned the model parameters---the finger moving mass $m$, damping $b$, and stiffness $k$---as dependent variables, the sliding speed and applied force as the main effects, as well as their interaction, and participant ID as a random effect. 

The GLMM analysis revealed that the finger moving mass was affected by sliding speed ($F(1,205) = 7.193$, $ p=0.008$) with a fixed coefficient of $3.5 \times 10^{-5}$~Kg/mm/s, but not by the applied force ($F(1,176) = 0.125$, $p=0.724$) or their interaction ($F(1,201) = 1.216$, $p=0.271$). The estimated coefficient indicates an increase of approximately $0.0028$~kg in the moving mass from the lowest to the highest sliding speed. Normalizing this value with respect to the moving mass from the lowest speed, we observed a 0.82\% increase in moving mass for a unit increase in sliding speed. 

The finger damping was not affected by the applied force ($F(1,179) = 3.280$, $p=0.071$), sliding speed ($F(1,224) = 2.843$, $p=0.093$), or their interaction ($F(1,221) = 1.604$, $p=0.207$). 

The finger stiffness was affected by sliding speed ($F(1,208) = 14.465$, $p<0.001$) with a fixed coefficient of $4.036$~N/m/mm/s, but not by the applied force ($F(1,175) = 1.148$, $p=0.285$) or their interaction ($F(1,204) = 3.79$, $p=0.053$). Based on the obtained coefficient values, the finger stiffness increased by approximately $322.88$~N/m from the lowest to the highest sliding speed. When normalized relative to the stiffness at the lowest speed, this value corresponds to a 1.11\% increase in stiffness per unit increase in sliding speed. 

Moreover, the obtained moving mass ($-2LL_{reduced} = -2267.563
$ and $-2LL_{full}= -2352.209$, yielding an $LRT$ = 84.646 $>$ 3.84), damping ($-2LL_{reduced} = 592.744
$ and $-2LL_{full}= 550.981$, yielding an $LRT$ = 41.763 $>$ 3.84), and stiffness ($-2LL_{reduced} = 3301.393
$ and $-2LL_{full}= 3241.765$, yielding an $LRT$ = 59.628 $>$ 3.84) values were significantly affected by the participant ID. Additionally, a post-hoc Wilcoxon signed-rank test was performed to determine which sliding speed conditions produced statistically significant differences in these skin mechanics parameters. Significant comparisons are indicated by the bars above the data in Figures~\ref{fig: mech_mass},~\ref{fig: mech_damp}, and~\ref{fig: mech_stiff}. 

\begin{figure}[!t]
     \centering
     \includegraphics[width=0.8\linewidth]{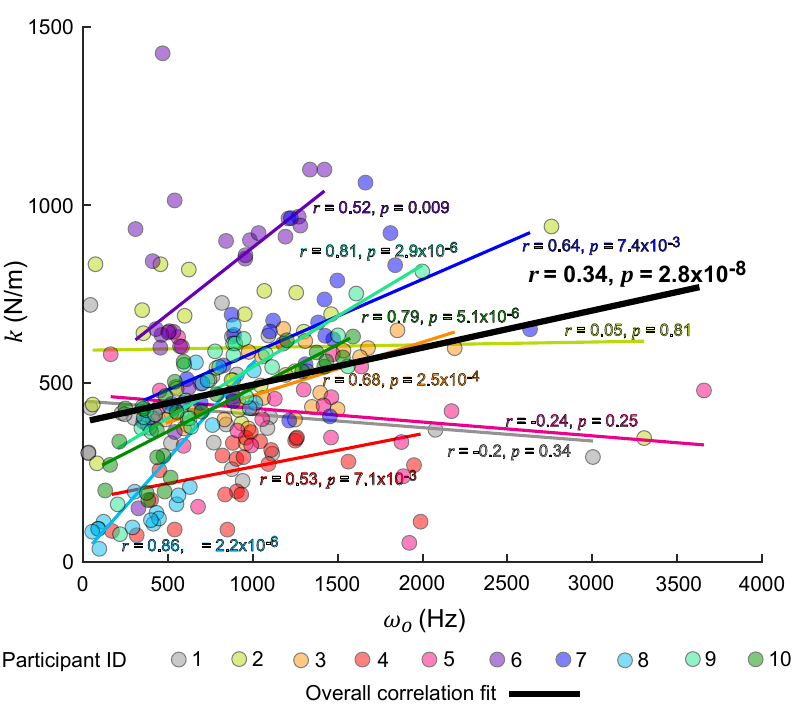}
     \caption{Scatter plot showing the Spearman's correlation between the cutoff frequency, $\omega_{o}$, of the electrovibration-induced finger friction response and finger stiffness, $k$. Colored dots represent individual data points for each participant, with corresponding dashed lines indicating participant-specific correlation fit. The Spearman correlation coefficient, $r$, and the associated significance value, $p$, are displayed on the plot. The overall correlation across all participants is represented by the solid black line.}
     \label{fig: corr_plot_wk}
 \end{figure}

\subsection{Percentage contribution of applied force and sliding speed}

Based on the GLMM coefficients, we computed the percentage contributions of applied force and sliding speed at preferred exploration conditions of $F_n = 0.4$~N and $\nu=80$~mms/s~\cite{kejriwal2023user} to the electrovibration-induced finger friction parameters ($K$ and $\omega_o$) and skin mechanics parameters ($m$, $b$, and $k$) as summarized in Table~\ref{tab: contributions}. For details regarding the calculation procedure, please refer to \ref{sec: percent_cal}.

\begin{table}[!ht]
\caption{Percentage contribution of finger-electrovibration interaction friction response parameters and skin's mechanical parameters.}
\centering
\begin{tabular}{cc|c|c|c|c}
\multicolumn{2}{c|}{Parameter}                                                                                                                      & \begin{tabular}[c]{@{}c@{}}Force \\ coefficient\end{tabular} & \begin{tabular}[c]{@{}c@{}}Speed\\ coefficient\end{tabular} & \%Force & \%Speed \\ \hline
\multirow{2}{*}{\begin{tabular}[c]{@{}c@{}}Electrovibration-induced\\  finger friction \\ response\end{tabular}} & Gain, $K$                                                              & -0.005                                                       & -4.581 $\times 10^{-5}$    & 99.1    & 0.9     \\
                                                                          & \begin{tabular}[c]{@{}c@{}}Cut off \\ frequency , $\omega$\end{tabular} & 1.121                                                        & 25.068                                                      & 18.85   & 87.35   \\ \hline
\multirow{3}{*}{Skin mechanics}                                           & Mass, $m$                                                               & 0.125                                                        & 7.193                                                       & 12.64   & 87.35   \\
                                                                          & Damping , $b$                                                            & 3.280                                                        & 2.843                                                       & 83.02   & 16.97   \\
                                                                          & Stiffness , $k$                                                           & 1.148                                                        & 14.465                                                      & 31.22   & 68.77  
\end{tabular}
\label{tab: contributions}
\end{table}

\subsection{Influence of skin mechanics on finger-display interaction}

Finally, we performed a Spearman's correlation~\cite{mukaka2012guide} analysis between the finger-display interaction parameters (the gain, $K$, and cutoff frequency, $\omega_o$) and the finger mechanical response parameters (the moving mass, $m$, damping, $b$, and stiffness, $k$) across all sliding speeds and force, as shown in Figure~\ref{fig: corr_plot}. The results indicate that the gain decreased with increasing finger moving mass, and increased with increasing finger damping. Similarly, the cutoff frequency increased with higher finger moving mass. Among all parameter pairs, only the cutoff frequency and finger stiffness exhibited a low positive correlation ($r = 0.34, p = 2.8 \times 10^{-8}$, see Figure~\ref{fig: corr_plot_wk}). Furthermore, seven out of ten participants showed a correlation greater than 0.52 ($p < 0.009$)between their finger stiffness and cutoff frequency.

\section{Discussion}\label{sec:discussion}

In this study, we investigated how exploration conditions influence electrovibration-induced finger friction on touchscreens and the role of skin mechanics in this relation, with particular focus on texture rendering. We measured the contact forces and skin accelerations generated as ten participants moved their index fingers on an electrovibration display under varying exploratory forces and speeds. Using this data, we developed mathematical models to characterize the electrovibration-induced finger friction and the finger vibrations it causes. Finally, we analyzed how the models' parameters change across different sliding speeds and applied force settings.

Our results demonstrate that the cutoff frequency of electrovibration-induced finger friction response increased with sliding speed (Figures~\ref{fig: elecmech} and \ref{fig: elecmech_cutoff}). We attribute this increase to changes in skin mechanics with sliding speed. In our measurements, both the finger moving mass and stiffness increased with sliding speed and consequently influenced the natural frequency of the finger dynamics, $\omega_{finger} = \sqrt{{k}/{m}}$; check Figure~\ref{fig: mech_mass} and~\ref{fig: mech_stiff}. While higher stiffness raises the resonance frequency, the increased moving mass lowers it, potentially reducing the cutoff frequency. However, our GLMM analysis revealed that the variation in the stiffness was stronger than the change in mass, resulting in an overall increase in the cutoff frequency. For every 1~mm/s increase in sliding speed, the predicted finger stiffness increased by 1.11\% while the finger moving mass increased by 0.82\%. 
Furthermore, we observed statistically significant correlations between estimated cutoff frequency and finger stiffness (Figure~\ref{fig: corr_plot_wk}), reinforcing the role of stiffness as the dominant factor. These findings indicate that with increasing speed, the finger exhibits less elastic behavior, causing an increase in the cutoff frequency. 

The increase in finger stiffness with sliding speed can be attributed to additional shear forces generated during hand motion~\cite{shao2016spatial,adams2013finger}. When an alternating voltage is applied to the electrovibration touchscreen, the finger is repeatedly attracted to and released from the surface. As the voltage approaches zero, the attraction force diminishes, allowing the shear strain to pull the finger back---a phenomenon that intensifies with increasing finger speed~\cite{vardar2017effect}. As the shear strain amplitude increases, finger stiffness increases, decreasing its elasticity~\cite{pataky2005viscoelastic, wang2007vivo,delhaye2016surface}. As a result, less energy is dissipated at higher frequencies, as more vibration energy is transmitted into the bulk of the finger~\cite{serina1997force}. Additionally, amplitude-modulated input further facilitates the vibration energy transmission~\cite{lamore1986envelope} to the finger bulk, collectively contributing to an increase in the cutoff frequency of the finger's electrovibration friction response. 

Unlike the cutoff frequency, our results (Figures~\ref{fig: elecmech} and ~\ref{fig: elecmech_gain}) showed that the gain of the electrovibration-induced finger friction was not affected by the sliding speed. 
We explain this invariance with the observed simultaneous variation in finger moving mass and damping. Our correlation results revealed that the gain decreased with increasing finger moving mass, but increased with increasing finger damping (see Figure~\ref{fig: corr_plot}(a) and Figure~\ref{fig: corr_plot}(c)). A higher finger moving mass implies a lower finger vibration amplitude at low frequencies, which is consistent with our results. Although the finger moving mass increased with speed, the increase is $3.223\times 10^{-5}$~kg per mm/s speed, which is small enough that it does not drastically affect the gain. Similar to the finger moving mass, an increase in finger damping reduces the finger vibration intensity at the resonance. However, our results showed that, like the gain, finger viscosity also remained constant across the sliding speed. 

Our results showed that applied force did not have a significant effect on the electrovibration-induced friction and on the finger vibration caused by it (see Figures,~\ref{fig: elecmech},~\ref{fig: elecmech_gain},~\ref{fig: elecmech_cutoff},~\ref{fig: mech_mass},~\ref{fig: mech_damp}, and ~\ref{fig: mech_stiff}). Interestingly, previous studies have reported an increase in generated friction force due to electrovibration with applied force~\cite{guo2019effect}. Similarly, earlier studies~\cite{hajian1997identification, kao2004stiffness, wiertlewski2012mechanical} observed a rise in the finger's mechanical parameters, such as mass, damping, and stiffness with applied force. This contradiction between our results and earlier studies could be caused by the low force ranges for the active applied forces ($0.2 \sim 0.6$~N) in our study ~\cite{papetti2016vibrotactile}. Vardar and Kuchenbecker~\cite{vardar2021finger} also found no significant change in the relationship between the input voltage and electrovibration force with varying applied force. They argued that normal forces at 0.5 and 1.0~N still represent sufficiently light and moderate touches not to cause any change in the skin mechanics, which is consistent with our results. 

Our findings revealed substantial inter-participant variability in electrovibration-induced finger friction (Figures~\ref{fig: elecmech},~\ref{fig: elecmech_gain}, and~\ref{fig: elecmech_cutoff}) and skin mechanics (Figures~\ref{fig: mech_mass}, \ref{fig: mech_damp}, and \ref{fig: mech_stiff}). These differences were statistically significant, as confirmed by likelihood ratio tests on the GLMM models, which showed that the electrovibration-induced finger friction response parameters ($K$ and $\omega_o$) and skin mechanics parameters ($m$, $b$, and $k$) significantly varied between participants. 
Similar inter-participant variability has been reported in previous studies on the finger-electrovibration interaction studies ~\cite{vardar2017effect, guo2019effect, vuik2024impact}. Such variability is likely driven by individual differences in finger and skin biomechanics, skin hydration, and finger size~\cite{woodward1993relationship,neely2006gender,wiertlewski2012mechanical,adams2013finger,park2014mechanical,abdouni2017biophysical, abdouni2018impact, infante2025role}. These findings suggest that the design of haptic feedback in electrovibration displays may benefit from accounting for individual differences in skin properties.

Our results indicate that at low applied forces ($0.2$~N to $0.6$~N), the spectrum for the electrovibration-induced finger friction remains unchanged. However, as sliding speed increases, the higher-frequency signals of the friction spectrum experience less attenuation, making these frequencies more pronounced. If this speed-dependent behavior is not accounted for in electrovibration-based texture rendering, users may perceive virtual textures differently from the intended ones (e.g., finger friction between real textures or a designed virtual slider in a car cockpit). To ensure a more realistic and robust experience, texture rendering algorithms should dynamically adjust any high-frequency signal based on the sliding speed. 

Based on these findings, we proposed an empirically derived model (Equation~\eqref{eqn: finger_display_model}) that predicts the evolution of finger friction generated by electrovibration based on finger sliding speed for the given input signal magnitude. This model enables researchers to implement a corresponding inverse filter within texture rendering algorithms to minimize the discrepancy between the intended texture spectrum and the actual friction spectrum. The filter dynamically adjusts input voltage to produce the desired frictional response during finger interaction with electrovibration surfaces. In future work, we plan to integrate this filter into a closed-loop texture rendering algorithm that adapts in real time to user exploration, supporting a more natural and freeform tactile interaction experience. 

Notably, our model (Equation~\eqref{eqn: finger_display_model}) is based on parameters averaged across participants, following a methodology similar to that of Felicetti et al.~\cite{felicetti2023investigation}, who developed a generalized model by averaging the transfer functions of input voltage to output vibration across users in a vibrotactile device for texture rendering. We plan to evaluate the perceptual consequences of using averaged electrovibration-induced finger friction response versus participant- or condition-specific compensations.  

Despite our careful investigation, our work still has limitations. Participants were instructed to perform a constrained active finger exploration by following the cues in the experimental GUI. However, this task proved difficult for some individuals, resulting in the exclusion of data from seven participants due to inconsistencies or errors in their task execution. Participants reported difficulty maintaining a consistent contact angle, applied force, and sliding speed during exploration. Such variations likely influenced the finger-contact behavior, thereby affecting the measured interaction forces~\cite{vardar2021finger}. The experiments were also conducted over multiple sessions, which may have introduced further variability in the estimated model parameters. This variability is further reflected in the large standard errors observed in Figure~\ref{fig: elecmech}, particularly at higher applied forces and lower sliding speeds, indicating difficulties in maintaining consistent exploration behavior. Although the trial rejection method described in \ref{sec:AppC} helped ensure data reliability, a more controlled approach, such as passive finger exploration, could improve consistency. In future work, we plan to address this limitation by developing a motorized linear-stage setup to standardize finger exploration and enable data collection from a broader participant population.

Moreover, we employed simplified models to represent electrovibration-induced finger friction (a first-order system model) and skin's mechanical response (a second-order mass-spring-damper model), treating them separately. While higher-order models that integrate both responses are feasible, we deliberately adopted this simplified approach for texture rendering applications on electrovibration displays. As previously discussed, texture rendering on electrovibration displays relies on modulating friction by applying voltage signals, which must be calibrated to match the exploration condition for robust results, as evidenced by our results. Using a higher-order models to represent the full interaction dynamics could lead to overcompensation during real-time rendering. Additionally, to better understand the effect of skin mechanics on the estimated finger-display interaction parameters, we opted for a one-degree-of-freedom (DOF) second-order model. While a multi-DOF model would more accurately capture the complexity of skin behavior, its implementation would require measurements of vibration velocities at multiple locations on the finger, such as the contact point and the nearby regions, accounting for the motion of independently vibrating mechanical elements. However, such a detailed understanding of internal dynamics may not necessarily provide additional insight into how model parameters influence the finger-display interaction response, nor significantly enhance rendering fidelity. 

Nevertheless, we acknowledge that the simplified nature of our models may overlook complex interfacial phenomena, such as additional damping and stiffness effects, which are important for advancing the scientific understanding of finger-surface interactions. As part of future work, we aim to develop a more comprehensive multi-DOF model that more accurately captures these dynamics. For this goal, we will measure multiple independent vibration vectors between the contact point and the finger nail, enabling a 2-DOF or multi-DOF representation. Currently, our skin mechanics model is limited by its reliance on a single vibration signal, measured as acceleration at the fingernail. 

\section{Conclusions}
Our results demonstrate that, within the sliding range of 20--100~mm/s, finger stiffness increases by 4.04~N/m for each unit increase in speed (relative change of 1.11\%), which exceeds the corresponding increase in moving mass, measured at $3.23\times10^{-5}$~kg (0.82\%). This disparity results in a 13.8~Hz (2.47\%) rise in the cutoff frequency of the electrovibration-induced finger friction response. Furthermore, we found that low to moderate applied forces (approximately 0.1~N to 0.6~N) had no statistically significant effect on either the skin's mechanical properties or the parameters of the electrovibration-induced finger friction response. Based on these findings, we developed a practical, speed-dependent model that dynamically adjusts the input voltage to achieve the desired output friction. This model enables researchers to enhance the realism of texture rendering algorithms by minimizing discrepancies between the target texture spectrum and the resulting electrovibration-induced finger friction spectrum, thereby supporting more consistent and perceptually accurate haptic feedback. 

\section{Data availability}
The data and code used in this paper will be publicly available upon acceptance. 
\section{Acknowledgements}
The authors thank Telesilla Bristogianni for helping us with cutting the 3M touchscreen. The authors also thank Michaël Wiertlewski for lending the force sensor for our experimental setup and Celal Umut Kenanoğlu for his feedback on the results and discussion sections.
\section{Declaration of competing interest}
The authors declare they have no known competing financial interests or personal relationships that could have influenced this work.
\section{Declaration of generative AI and AI-assisted technologies in the writing process}
The authors used OpenAI’s ChatGPT and Grammarly Inc.’s Grammarly tool to enhance the language and read-
ability of this work. After using these tools, the authors carefully reviewed and edited the content as needed. They
take full responsibility for the final publication.
\section{Funding statement}
This work was partly funded by the Dutch Research Council (NWO) with project number 19153.
\section{CRediT authorship contribution statement}
\noindent\textbf{Jagan K. Balasubramanian:} Conceptualization, Methodology, Software, Formal Analysis, Investigation, Writing - Original Draft, and Visualization.\\
\textbf{Daan M. Pool:} Methodology, Formal Analysis, Writing - Review \& Editing.\\
\textbf{Yasemin Vardar:} Conceptualization, Methodology, Formal Analysis, Writing - Original Draft,  Writing - Review \& Editing, Supervision, and Funding Acquisition.

\appendix
\section{Electrovibration force generation} \label{sec:appA}

\subsection{Voltage doubling}
This section explains how the input message signal, $V_m(t)$, causes electrostatic force $F_e(t)$ at twice the message signal's frequency. For simplicity, let us assume the input is a pure sine wave with frequency $ f_m$.

\begin{equation}
    V_m(t) = V_m \cos(2\pi f_m t)
\end{equation}

From Vardar et al~\cite{vardar2021finger}, we know $F_e \propto kV_m^2$, where $k$ is a constant. Substituting this relation with the previous equation, we get
\begin{equation}
   F_e(t) \propto k (V_m \cos(2 \pi f_m t))^2
\end{equation}
\begin{equation}
   F_e(t) \propto k V_m^2\frac{1+\cos(2 \times 2 \pi f_m t)}{2}
\end{equation}

The equation above shows that the generated electrostatic force has twice the frequency of the input signal.

\subsection{Generated friction force through amplitude modulated input signal in electrovibration}

This section explains how an electrovibration stimulus generates finger friction through an amplitude-modulated input signal. We first take the square root of the input voltage signal, $V_m(t)$, to remove the nonlinearity. Additionally, we DC shift the input voltage signal with $\min|V_m(t)|$ to ensure the amplitude is real. The envelope for the amplitude-modulated signal is $V_{env}(t)$ is given as:

\begin{equation}
    V_{env}(t) = \sqrt{V_m(t)+\min|V_m(t)|}
\end{equation}

For amplitude modulation, we multiply the above equation by a high-frequency carrier wave, $f_c$, usually beyond 5~kHz. The DC shift, $\min|V_m(t)|$, causes the amplitude modulation to be a double-sideband full carrier. The peak amplitude of $V_m(t)$ and carrier is 1V, resulting in a modulation index of 1. The modulation index should be $<=1$ to prevent signal distortion due to over-modulation. We chose the modulation index as unity to ensure the input message signal's power is higher. The carrier signal does not contain useful information for texture perception since the electrovibration friction intensity depends on the input voltage signal's, $V_i(t)$, power.

\begin{equation}
    V_{i}(t) = \sqrt{V_m(t)+\min|V_{m}(t)|} \cos(2 \pi f_c t)
\end{equation}

The generated electrovibration force and the input voltage are related through the following equation. 

\begin{equation}
    F_e = k V_i(t)^2,
\end{equation}

where, $F_e$ is the electrovibration force, and $k$ is a proportionality constant .

\begin{equation}
    F_e = k \sqrt{V_m(t)+\min|V_m(t)|} \cos(2 \pi f_c t))^2
\end{equation}

\begin{equation}
    F_e = k (V_m(t)+\min|V_m(t)| )\frac{1 + \cos(2 \pi f_c t)}{2} 
\end{equation}

\begin{equation}
    F_e = \frac{k}{2}  (V_m(t)+\min|V_m(t)| + (V_m(t)+\min|V(t)|)\cos(2 \pi f_c t))
\end{equation}

\begin{equation}
    F_e = \frac{k}{2}  (V_m(t)+\min|V_m(t)| + V_m(t)\cos(2 \pi f_c t)+\min|V_m(t)|\cos(2 \pi f_c t))
\end{equation}

Substituting $V_m(t)=V_mcos(2\pi f_mt)$ in the above equation.

\begin{equation}
    \begin{aligned}
        F_e = \frac{k}{2}  (V_m\cos(2 \pi f_m t)+\min|V_m(t)| + V_m\cos(2 \pi f_m t)\cos(2 \pi f_c t)+ \\ \min|V_m(t)|\cos(2 \pi f_c t))
    \end{aligned}
\end{equation}

\begin{equation}
    \begin{aligned}
          F_e = \frac{k}{2}  (V_m\cos(2 \pi f_m t)+\min|V_m(t)| + V_m(\cos(2 \pi f_c + f_m t) + \\ \cos(2 \pi f_c-f_m t))+\min|V(t)|\cos(2 \pi f_c t))  
    \end{aligned}
\end{equation}

Since the finger acts as a low-pass filter with a cutoff at 1~kHz, the higher-frequency components are attenuated:

\begin{equation}
    \begin{aligned}
            F_e = \frac{k}{2}  (V_m\cos(2 \pi f_m t)+\min|V_m(t)| + \cancel{V_m(\cos(2 \pi f_c + f_m t) + \cos(2 \pi f_c-f_m t))} \\ + \cancel{\min|V_m(t)|\cos(2 \pi f_c t))}
    \end{aligned}
\end{equation}

The resulting friction force, $f$, from the electrostatic force, $F_e$ and the finger normal force, $F_n$, is expressed as:

\begin{equation}
    f = \mu (F_e + F_n) = \mu (\frac{k}{2}  (A_m\cos(2 \pi f_m t)+\min|V_m(t)|) + F_n),
\end{equation}

Where $\mu$ is the friction coefficient.

\section{Setup response} \label{sec:AppB}

We wanted to ensure that the experimental setup has a flat response till 1000~Hz as the human tactile range is from 1~Hz to 1000~Hz~\cite{bolanowski1988four}. We used an impact hammer to excite the system in the normal and lateral directions to measure the response of the experimental setup. The response along the normal direction was found by deriving the transfer function of the setup with vibration as velocity signal, $v_{snd}$, from the impact hammer as output, and the input was recorded from the force sensor, $F_{snd}$. We used $v_{snd}$ as input and $F_{snd}$ as output to make the transfer function causal. In the normal direction, we observed that the setup had a flat response till 500~Hz and a resonance at 1.4~kHz. We fit a second-order model to the setup response in the normal direction, $H(s)_{snd}$ defined as 

\begin{figure}[!ht]
    \centering
    \includegraphics[width=0.55\linewidth]{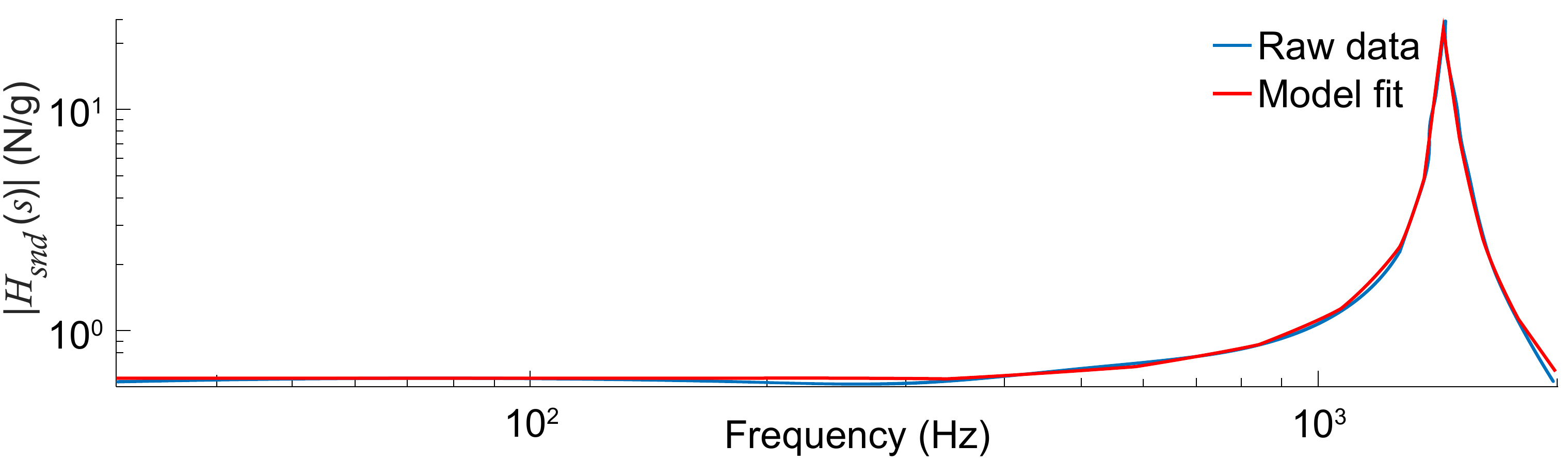}
    \caption{Frequency response of the experimental setup measured in the normal direction. The setup was excited with an impact hammer. The output response along the normal direction was measured using a force sensor.}
    \label{fig: Normal_response}
\end{figure}
\begin{figure}[!ht]
    \centering
    \includegraphics[width=0.55\linewidth]{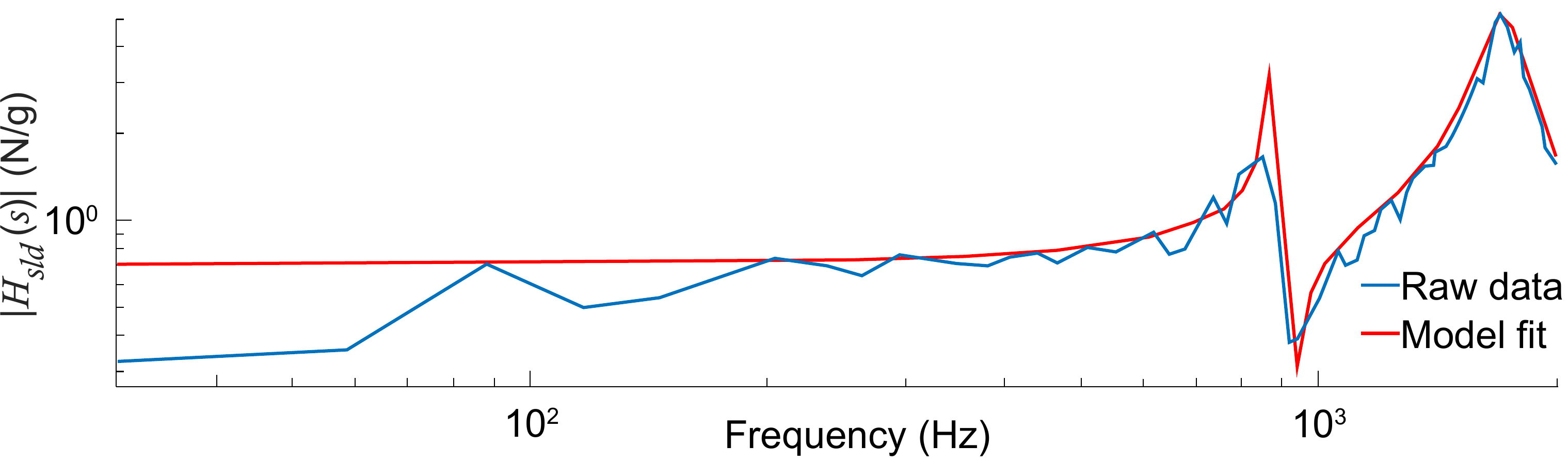}
    \caption{Response of the setup measured in the lateral direction. The output response was measured along the force sensor's lateral axis.
    }
    \label{fig: Lateral_response}
\end{figure}

\begin{equation}
\frac{v_{snd}(s)}{F_{snd}(s)}=H_{snd}(s) = \frac{sK_{snd} \omega_{n, snd}^2}{s^2+2\zeta_{snd} \omega_{n, snd} s+\omega_{n, snd}^2}
\end{equation}

Here, $K_{snd}$ is the gain, $\omega_{n, snd}$ is the natural frequency, and $\zeta_{snd}$ is the damping coefficient along the normal direction. By fitting the model to the raw data, we determined the parameters as $K_{snd} = 0.58$, $\omega_{n, snd} = 1454$~Hz, $\zeta_{snd} = 0.011$. We believe the double-sided tape attenuated the measured force, which we accounted for with $K_{snd}$. When measuring the force sensor output without the touchscreen and tape, the magnitude was $10^0$. 

In the lateral direction, we measured two resonances at 866~Hz and 1740~Hz. We think that the use of double-sided mounting tape caused the first resonance. When measuring the force sensor’s response without the touchscreen and the tape, we detected a single resonance at 2.4~kHz, consistent with the sensor’s datasheet. Hence, we modeled the setup as a fourth-order system, incorporating the mounting effects, as described by the following equation:
\begin{equation}
   \frac{v_{sld}(s)}{F_{sld}(s)} = H_{sld}(s) = \frac{s\omega_{n, sld}^2}{s^2+2\zeta_{sld} \omega_{n, sld} s+\omega_{n, sld}^2}\cdot\frac{b_3s^2+b_2s+b_1}{a_3s^2+a_2s+a_1}
    \label{eqn: setup response}
\end{equation}
Here, $v_{sld}(s)$ and $F_{sld}(s)$ are the Fourier transforms of the velocity signal from the impact hammer and lateral force signal from the force sensor, $\zeta_{sld}$ represents the damping coefficient, and $\omega_{n, sld}$ denotes the natural frequency in the lateral direction. The mounting effects are modeled by the transfer function $\frac{b_3s^2 + b_2s + b_1}{a_3s^2 + a_2s + a_1}$.  Figure~\ref{fig: Lateral_response} illustrates the system's lateral response. The measured values were $ \omega_{n, sld} = 1714$~Hz, $\zeta_{sld} = 0.06$, $b_3 = 1$, $b_2 = 92$, $b_1 = (2\pi\cdot922)^2$, $a_3 = 1$, $a_2 = 0.0001$, and $a_1 = (2\pi\cdot856)^2$. 

\subsection{Removing setup response}

We noticed that the input message signal to output friction raw response and the setup's response had a first-order behavior, as shown by the sweep markers in Figure~\ref{fig: Model data}. The equation gives the response:

\begin{equation}
    \frac{f(s)}{V_m(s)} = H_{f-e}(s) \cdot H_{sld}(s)
    \label{eqn: Combined response}
\end{equation}

Here, $f(s)$ and  $V_m(s)$ are the Fourier transforms of the friction signal from the force sensor and the input voltage envelope. $H_{f-e}(s)$ is the transfer function of finger electrovibration interaction and $H_{sld}$ is the transfer function of the setup's response. After removing the setup's response, $H_{sld}$, from the Equation~\eqref{eqn: Combined response}, we get a first-order response for $H_{f-e}$ which is modeled as:

\begin{equation}
    \frac{f(s)}{V_m(s)} = H_{f-e}(s) = \frac{K\omega_o}{s+\omega_o}
\end{equation}

Here $K$ is the system gain, and $\omega_o$ is the cutoff frequency.

\section{Exploration condition check} \label{sec:AppC}

During the experiment, we ensured that participants adhered to the specified exploration conditions in each trial. To verify this, we recorded both the applied force and the scanning speed of the participant's finger throughout each trial. Applied force data were sampled at 20~kHz, and scanning speed at 60~Hz, resulting in 200,000 and 600 samples over a 10-second trial. Threshold ranges were defined as $\pm10\%$ for applied force and $\pm25\%$ for sliding speed, relative to target values.

\begin{figure}[!h]
    \centering
    \includegraphics[width=0.6\linewidth]{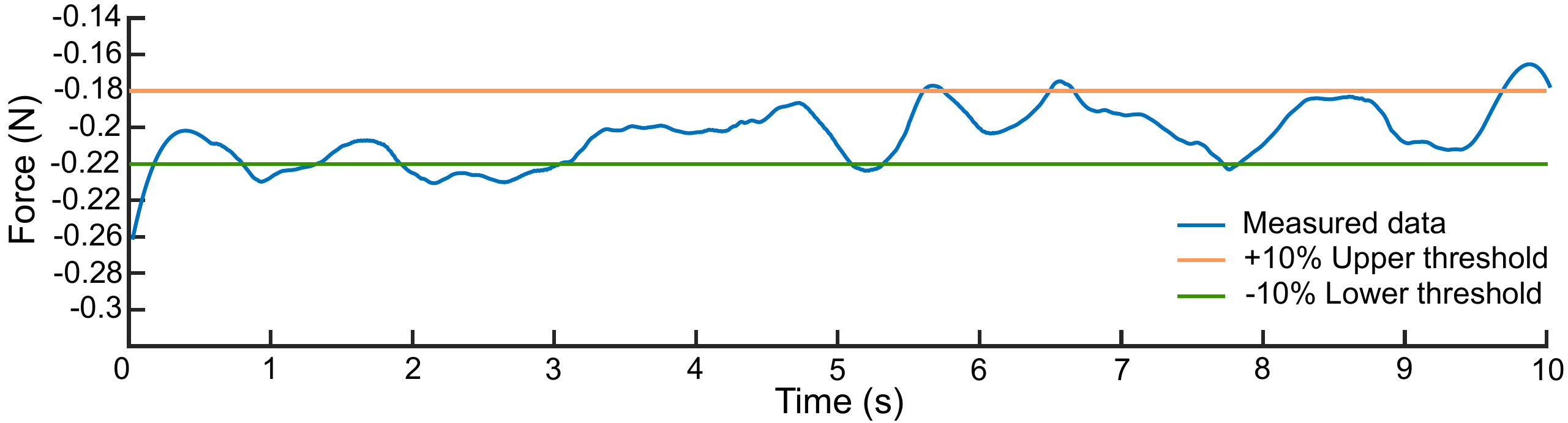}
    \caption{Illustration of force applied by a participant in a trial for the 0.2~N target setting.}
    \label{fig: Force_check}
\end{figure}

\begin{figure}[!ht]
    \centering
    \includegraphics[width=0.6\linewidth]{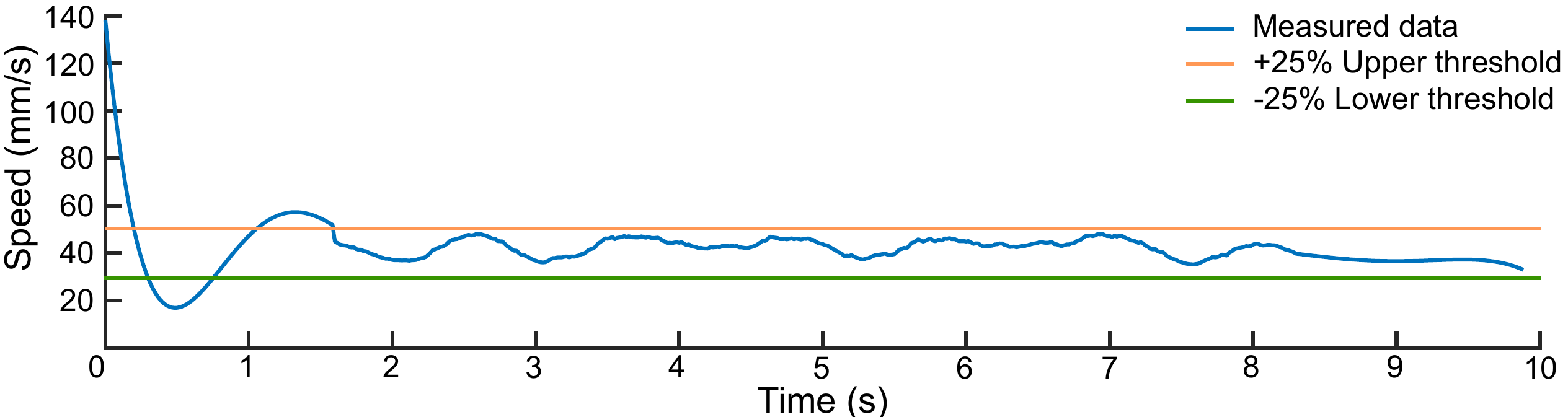}
    \caption{An illustration of the finger scanning speed recorded during a trial conducted with a 40~mm/s target condition.}
    \label{fig: Speed_check}
\end{figure}

For each trial, we first counted the number of samples that exceeded the $\pm25\%$ threshold for each parameter. Additionally, we computed the mean applied force and sliding speed and verified whether these values remained within $\pm10\%$ of the corresponding targets. A trial was deemed invalid and repeated if either of the following conditions was met: 1) the number of out-of-threshold samples exceeded the equivalent of one second of data---i.e., more than 20,000 samples for applied force or more than 60 samples for scanning speed---or 2) the mean value fell outside the $\pm10\%$ range. Figures~\ref{fig: Force_check} and~\ref{fig: Speed_check} present representative data for applied force and sliding speed from a single trial, illustrating the verification process.

\section{Finger exploration trajectory}\label{AppE}

\begin{figure}[!ht]
    \centering
    \includegraphics[width=0.5\linewidth]{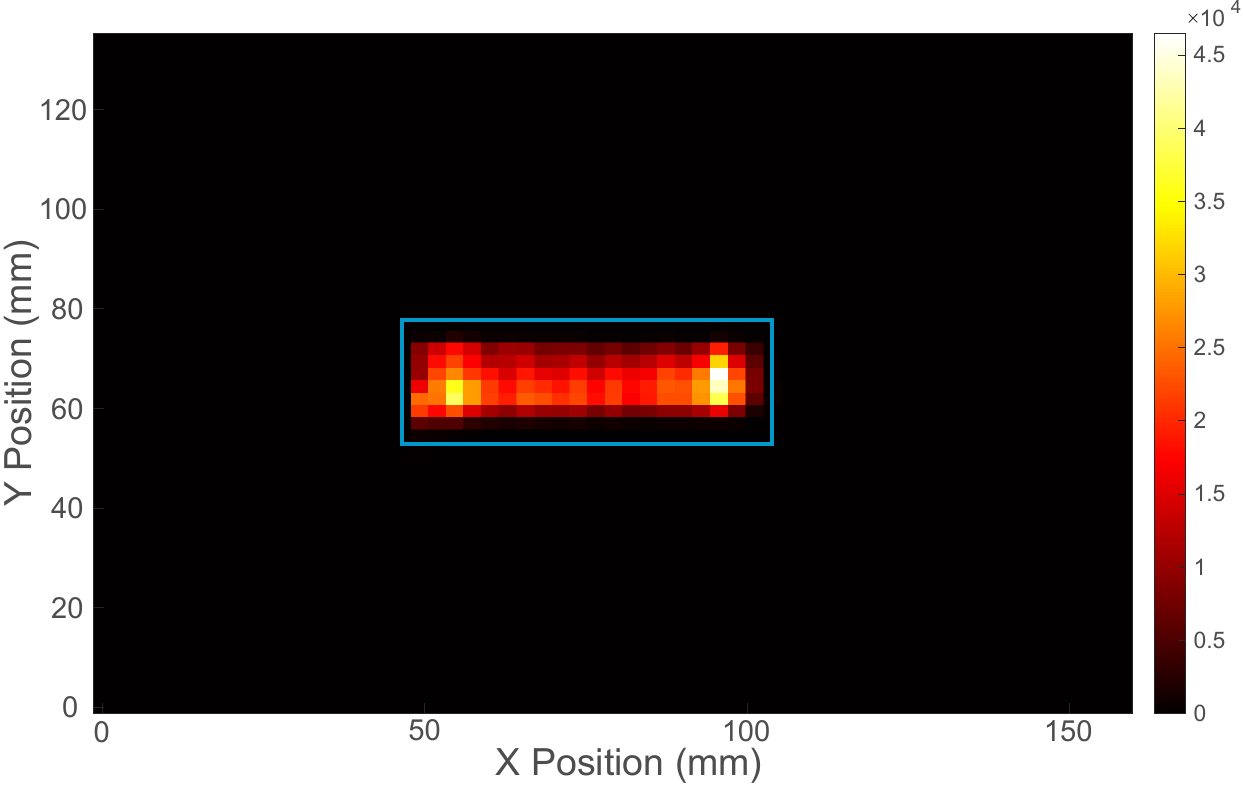}
    \caption{Heatmap of the finger exploration trajectory of all participants across all trials. The cyan box indicates the dimensions of the touchscreen.}
    \label{fig: Finger_trajectory}
\end{figure}

Participants used the index finger of their dominant hand to explore the touchscreen in lateral and medial directions. Figure~\ref{fig: Finger_trajectory} presents a heatmap of finger position data, sampled at 60~Hz across all participants and trials. The infrared position sensor covered an active area of 135~×~158~mm, with the 35~×~70~mm touchscreen centered within this area.

\section{Percentage contribution of applied force and scanning speed on GLMM model outputs}
 \label{sec: percent_cal}

Consider, for example, the equation for gain, $K$, modeled for fixed effects of applied force, $F_n$, and scanning speed, $v$, and corresponding coefficients, $\alpha_1$, $\alpha_2$, and $\alpha_3$ for applied force, scanning speed, and their interaction, respectively.

\begin{equation}
     K = \alpha_o+\alpha_1 F_n+\alpha_2 \nu+\alpha_3F_n*\nu+\rho+ \epsilon
\end{equation}

For the percentage contribution calculation, we compute the partial derivatives of the GLMM equation for $F_n$ and $v$ as $\frac{\partial K}{\partial F_n} = \alpha_1 + \alpha_3 v$ and $\frac{\partial K}{\partial v} = \alpha_2 + \alpha_3 F_n$.

Using the contribution of force,$C_F = \frac{\partial K}{\partial F_n} F_n$, and contribution of speed,$C_v = \frac{\partial K}{\partial v} v$, the percentage contributions are given as:

\begin{equation}
    \%Force = \frac{C_F}{C_F + C_V}\times 100
\end{equation}

\begin{equation}
    \%Speed = \frac{C_v}{C_F + C_V}\times 100
\end{equation}

The percentage contributions of each fixed effect on the GLMM outputs for the preferred applied force $F_n = 0.4$~N and scanning speed $v=80$~mms/s~\cite{kejriwal2023user} for $\alpha_1=-0.05$, $\alpha_2=-4.5 \times 10^{-5}$, and $\alpha_3 = 1.2 \times 10^{-4}$ are then $\%Force = 99.1$ and $\%Speed = 0.9$.
\newpage
\section{Spearman's correlation between finger-electrovibration response and skin mechanics} 

\begin{figure}[!h]
     \centering
     \includegraphics[width=0.7\linewidth]{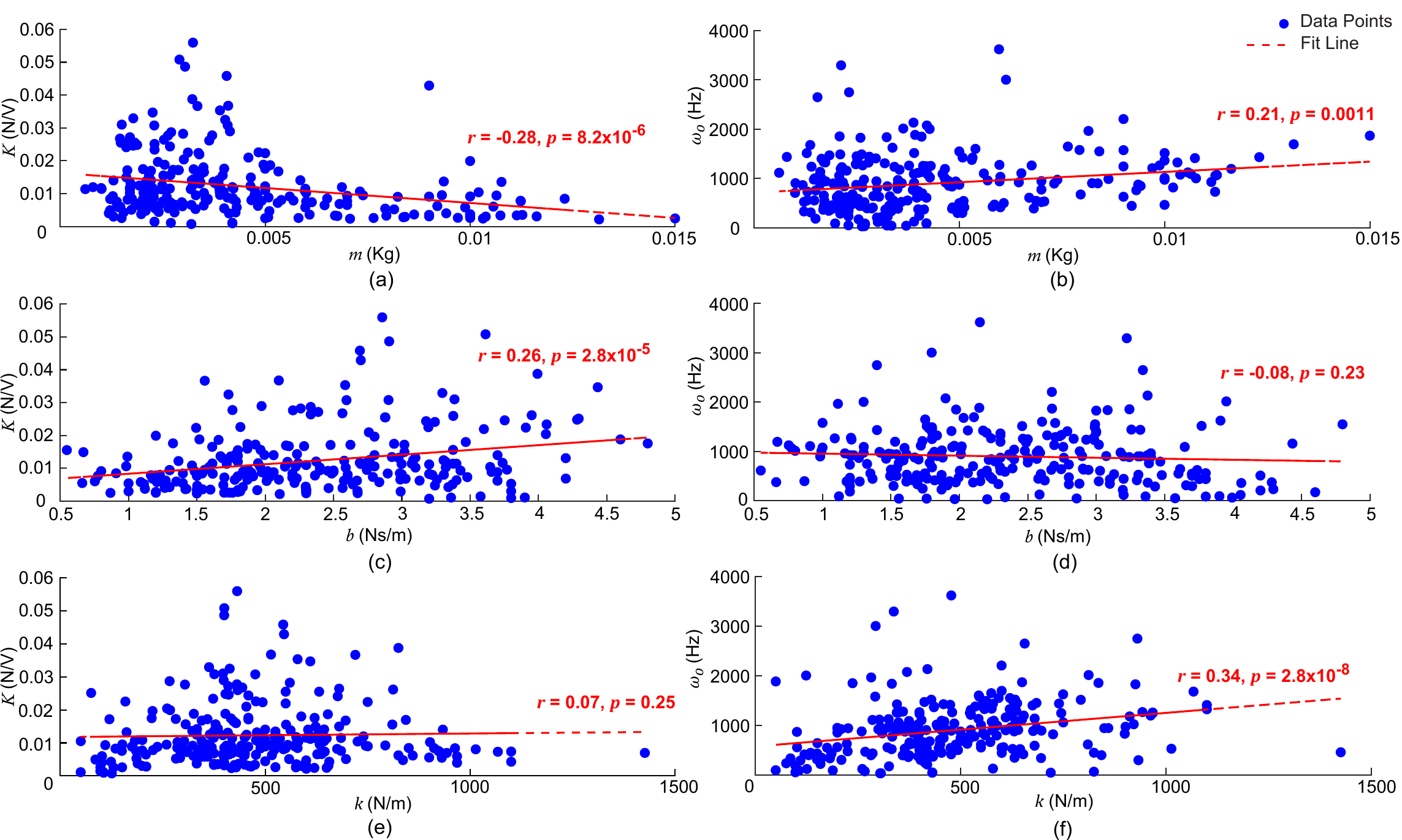}
     \caption{Scatter plot showing the Spearman's correlation between the finger-display interaction parameters---$K$ and $\omega_{o}$---and finger mechanical response parameters---$m$, $b$, $k$. (a) Gain, $K$ vs. finger moving mass, $m$, (b) cutoff frequency, $\omega_{o}$, vs. finger moving mass, $m$, (c) gain, $K$, vs. finger damping, $b$, (d) cutoff frequency, $\omega_{o}$, vs. finger damping, $b$, (e) gain, $K$, vs. finger stiffness, $k$, (f) cutoff frequency, $\omega_{o}$, vs. finger stiffness, $k$. The blue dots indicate the data points, and the dashed line indicates Pearson's correlation fit. The correlation coefficient, $r$, and significance value, $p$, are shown in the plot.}
     \label{fig: corr_plot}
 \end{figure}

\bibliographystyle{elsarticle-num-names}
\bibliography{ref}

\end{document}